\documentclass{revtex4}

\usepackage{graphicx}
\usepackage{color}

\begin{document}
\title{Low-energy spectrum of iron-sulfur clusters directly from  
many-particle quantum mechanics }

\author{Sandeep Sharma}
\affiliation{Department of Chemistry, Frick Laboratory, Princeton University, NJ 08544}
\author{Kantharuban Sivalingam}
\affiliation{Max Planck Institute for Chemical Energy Conversion, Muelheim, Germany}
\author{Frank Neese}
\affiliation{Max Planck Institute for Chemical Energy Conversion, Muelheim, Germany}
\author{Garnet Kin-Lic Chan}
\email{gkc1000@gmail.com}
\affiliation{Department of Chemistry, Frick Laboratory, Princeton University, NJ 08544}

\begin{abstract}
 FeS clusters are a universal biological motif. They carry out electron transfer, redox chemistry, and even oxygen
sensing, in diverse processes including  nitrogen fixation, respiration, and photosynthesis. 
The low-lying electronic states are key to their remarkable reactivity, but cannot 
be directly observed.
Here we present the first ever  quantum calculation of the
  electronic levels of [2Fe-2S] and [4Fe-4S] clusters free from any model assumptions.
  Our results highlight limitations of long-standing models of their electronic structure.
In particular, we  demonstrate that the widely used Heisenberg-Double-Exchange model underestimates the number of states by 1-2 orders of magnitude, which can conclusively be 
traced to the absence of Fe d$\rightarrow$d excitations,   thought to be important in these clusters. 
Further, the electronic energy levels of even the same spin are dense
on the scale of  vibrational fluctuations, and this provides a natural explanation for the ubiquity of these clusters  in nature for catalyzing reactions.
%% Our work illustrates the new 
%%   possibilities to realise bio-inorganic spectroscopy  from entangled many-particle quantum mechanical computations.
\end{abstract}

\maketitle

\newcommand{\fereduced}{{[Fe$_2$S$_2$(SCH$_3$)$_4$]$^{3-}$} }
\newcommand{\fedimer}{{[Fe$_2$S$_2$(SCH$_3$)$_4$]}}
\newcommand{\fecubane}{{[Fe$_4$S$_4$(SCH$_3$)$_4$]$^{2-}$}}

%by exploiting the area-law entanglement of the many body quantum states, 
%}%In particular, the widely used Heisenberg-Double-Exchange model completely misses the dense  spectrum  of the clusters, underestimating the number of low-lying states by 1-2 orders of magnitude.}
 %In the [4Fe-4S] cluster, 

Metals in enzymes perform remarkable chemistry under ambient pressures and temperatures. Among the most important cofactors are the iron-sulfur (FeS) clusters, 
comprising one to eight Fe atoms bridged by S ligands. In central processes of life ranging from nitrogen fixation to photosynthesis and respiration\cite{Beinert-science}, these
clusters perform diverse functions: redox chemistry, electron transfer, and even oxygen sensing.
Their electronic structure, with multiple low-lying states with differing electronic and magnetic character, 
holds the key to this rich chemistry. However, uncovering the electronic structure has been highly non-trivial. 
Direct experimental assignment of these electronic levels in  larger clusters has been impossible, since they lie at low energies and can be embedded 
within the vibrational modes of the clusters\cite{Johnson2005,Noodleman1995}. 

Nonetheless, through intense collaboration in the last decades between experiment and theory,
a consensus description of FeS cluster electronic states has emerged.
This is based on the Heisenberg-Double-Exchange (HDE) model, combining Heisenberg 
exchange between Fe spins with a simplified version of Anderson's double-exchange\cite{zener,anderson-DE} for 
mixed valence delocalization.
The HDE model was first proposed by Girerd\cite{girerd} and Noodleman et al.\cite{Noodleman1984,noodleman1986ligand} for [2Fe-2S]  dimers, and 
generalized to [3Fe-4S] and [4Fe-4S] clusters by Girerd, M\"unck and 
co-workers\cite{papa-fe3,blondin-90} with further extensions in recent years\cite{Borshch1984, guihery}.
The HDE model has yielded many important insights.
For example, the observed EPR $g<2$ in reduced [2Fe-2S] dimers was puzzling until it was 
recognized that the ground-state contains antiferromagnetic coupling of the 
Fe$^{\mathrm{II}}$ and Fe$^{\mathrm{III}}$ centers \cite{Gibson, Sands}. 
Similarly, in [3Fe-4S] and [4Fe-4S] clusters, HDE model mixed valence eigenstates provided the basis to interpret the distinctive M\"ossbauer, NMR, and 
ENDOR spectra\cite{Rius89,Mouesca1991,Bertini1991,Banci1991,Papa-mauss,Kappl2006}.

Despite the many successes of this phenomenological model, its limitations are also well-known. The HDE model 
posits couplings {\it a priori}, which must be fitted before making predictions. 
In basic versions, couplings consist of Heisenberg exchange $J$'s and double-exchange $B$'s; extended  versions include anisotropy, 
zero-field splitting, and further contributions\cite{Maurice2009}. An unambiguous determination of all parameters from experiment is clearly difficult, if not 
 impossible. For example, in one famous case of a [4Fe-4S] cluster\cite{Beinert-science}, the experimentally fitted $B$'s range from $\approx 10$ cm$^{-1}$ 
to $\approx 600$ cm$^{-1}$,  over two orders of magnitude! Further, FeS couplings obtained  
 from broken-symmetry density functional theory (BS-DFT)  computations (e.g. in pioneering work by Noodleman and others\cite{Noodleman1985,mouesca1994density,shoji2007theory,noodleman81,noodleman1986ligand,Yamaguchi1989210, Yamaguchi199056})
are not clearly more reliable. This is because BS-DFT only describes a weighted average over the (multiplet) states, and
individual parameters must once again be obtained through fitting;  further they depend strongly on the density functional approximation used\cite{Neese2009}.
Beyond the above issues, a deeper criticism of the phenomenological models is that they make {\it a priori} choices about relevant chemical processes:
 a pure Heisenberg model will completely  omit charge transfer phenomena. Without a reliable route to obtain model parameters, we cannot
verify their validity, which raises the troubling question: how can we be sure that the current models are even qualitatively correct?

\begin{figure}
\includegraphics[width=0.8\textwidth]{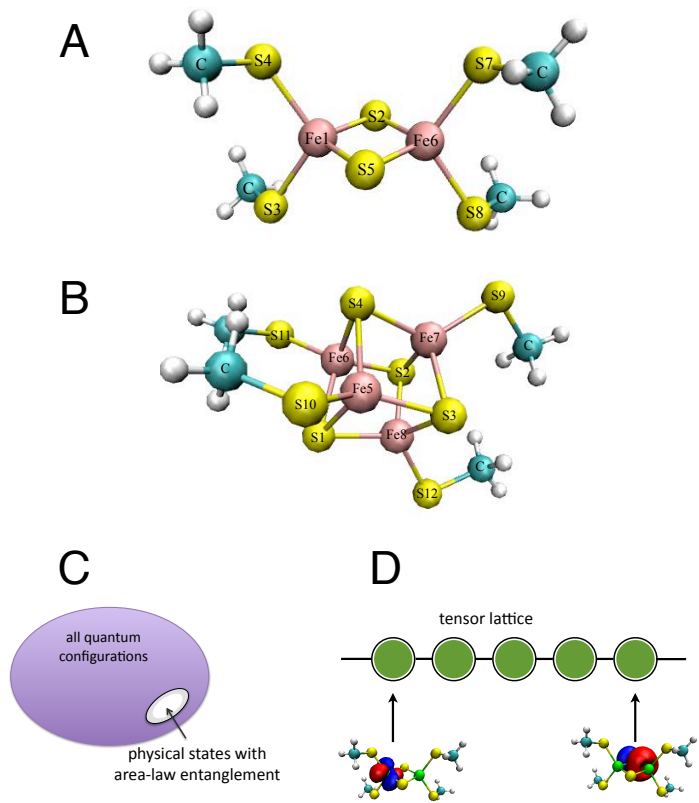}
\caption{{Model clusters and many-electron wavefunctions used in this work. 
%% he wavefunctions are represented by matrix product states (MPS) that exploit the underlying area-law entanglement of the low energy spectrum to simplify the calculations.
}
  (A) [2Fe-2S] and (B) [4Fe-4S] clusters in this work{, where the white circles denote H atoms. (Labels correspond to coordinates in Supplementary Tables 1, 2 and 11).} (C) The area-law entanglement of the physical states can be used to reduce the complexity of quantum calculations. (D) Wavefunctions with area-law entanglement can compactly be written as a tensor network where each tensor (represented here by a circle) denotes an active space orbital and the bonds between adjacent orbitals introduce local entanglement between them. The network shown here is the linear network used by the density matrix renormalization group, which removes the need to model unphysical entanglement between widely separated sites on the chain.
\label{fig:geom}}
\end{figure}

In principle, it should be possible to compute the states of the FeS clusters directly, without assuming an intermediate model.
After all, the long-standing premise of {\it ab-initio} computation is that any molecule's electronic structure is obtainable,
without assumptions, from the many-electron Schr\"odinger equation. Clearly, this has not yet been achieved for FeS clusters! 
Quantum mechanical computation is exponentially harder than classical simulation, because it
supports superposition and non-local correlations known as entanglement. To avoid this complexity, practical simulations - for example those using
DFT -  rely on a mean-field approximation, which  treats only classical-like quantum states without entanglement.
This approximation completely fails in FeS clusters, however, due to the Fe d-shells, which are near-degenerate on the Coulomb interaction scale (so-called strongly interacting), 
rendering the molecular 
orbital picture and concept of a single mean-field electronic configuration invalid. This is why BS-DFT calculations, the standard
computational method applied to FeS clusters , only provide an energetic average over the spectrum, and do not allow us
to directly obtain the individual electronic states.
 
Strongly interacting electronic structure {\it fundamentally} involves entangled superpositions of many valence configurations. 
Malrieu and co-workers\cite{Miralles1992, Miralles1993, Castell1999,Cabrero2000} showed that the main classes of configurations for 
 metal centers include the configurations involving metal d electrons (a complete active space or CAS), augmented with excitations of bridging ligand orbitals, or by including ligand
orbitals in the active space. However, even with these restrictions, the number of configurations grows exponentially with the number of 
atoms, and quickly becomes intractable. Early attempts to model these superpositions in
the [2Fe-2S] dimer had to make further drastic approximations,  completely removing 
non-bridging atoms, and including only d electrons in the active space~\cite{hubner}.
In a [4Fe-4S] cluster, the {\it minimal} CAS comprises all Fe 3d and S 3p valence shells, a distribution of 54 electrons in 36 spatial orbitals, 
or 10$^{16}$ configurations,  unmanageable on any computer!

Recent advances from quantum information and condensed matter theory demonstrate, however, that {\it physical} quantum states - the quantum states seen in Nature - 
are special and contain a hidden structure to their wavefunctions. In particular, the low energy states
possess area-law entanglement\cite{hastings},  reflecting the
locality  present in all physical systems regardless of interaction strength (Figure \ref{fig:geom}, panel C). This  structure implies that the coefficients of the valence configurations in
the FeS clusters are related in a special way, and the information compressed.
To encode this relationship, we write the wavefunction as a {\it tensor network}\cite{Verstraete2008}, 
of which the density matrix renormalization group (DMRG) of White\cite{White1992}, a linear tensor network, is the most widely used example. 
A tensor in a tensor network represents a local variational degree of freedom: 
in a molecule, a tensor might represent an atom or an orbital; and contraction of the tensors creates
the local entanglement, similar to a bond (Figure \ref{fig:geom}, panel D). The area-law implies
that if we restrict ourselves to the physically relevant sector of quantum states
 the tensors used to describe physical states can be of low-rank.
Working with low-rank tensor networks, we can  elevate quantum simulations from the mean-field level to the level of
the entangled quantum mechanics necessary to describe FeS clusters,  while significantly
ameliorating or in some cases completely bypassing the exponential complexity of the {\it general}  quantum mechanical formulation.

Starting with the work of White and Martin\cite{White1999}, our group and others have been developing tensor networks and the DMRG in the context of quantum chemistry\cite{chan2011,Kurashige2013,Sharma2012,Chan2002,Zgid2008spin,Moritz2005orb,kurashige,Legeza2003lif, marti}. 
Here, we show that using our {\it ab-initio} DMRG methodology, we can now numerically solve the valence many-particle quantum mechanical equations for
FeS clusters to  chemical precision. This allows us, for the first time, to directly compute and probe the individual states and spectra
 in FeS clusters as large as the [2Fe-2S] and [4Fe-4S] clusters. This direct computation  unshackles the discussion of FeS electronic structure
 from any earlier model assumptions. As we shall see, 
our calculations enable us to review, revisit and ultimately substantially revise, the historical models that have so far been the only basis for 
understanding electronic structure in these clusters, opening up the possibility to unlock the secrets of the chemistry of these clusters from direct simulation.

\section{Results and Discussion}
\subsection{[2Fe-2S] dimers.}

We first begin with [2Fe-2S] dimers. We consider the synthetic \fedimer$^{2-/3-}$ 
complexes\cite{Mayerle1975} that mimic the dimers prominently found in 
ferredoxins\cite{Orme-Johnson1973,holmsynthetic} (Figure \ref{fig:geom}, panel A). 

In canonical understanding based on the HDE model, the Fe atoms are placed in definite valence states. For the oxidized { [2Fe-2S]$^{2-}$ }dimer, both Fe atoms are assumed
high-spin Fe$^{\mathrm{III}}$ ($S_{1,2}$=5/2), while for the reduced {[2Fe-2S]$^{3-}$} dimer, one is high-spin Fe$^{\mathrm{II}}$ ($S_1$=2) and the other, high-spin 
Fe$^\mathrm{III}$ ($S_2$=5/2). In the oxidized dimer there is assumed to be no double-exchange, and the HDE model reduces to the simpler Heisenberg form,
\begin{equation}
  H= 2JS_1\cdot S_2 \label{eq:heis}
\end{equation}
with levels $E(S)= JS(S+1)$ ($S$ is total dimer spin, and $J$ is exchange coupling). In [2Fe-2S] dimers, 
$J>0$, thus the ground-state is low-spin with maximal antiferromagnetic alignment.

In the canonical picture of the reduced dimer, the additional electron can delocalize between the two Fe centers. 
(This is influenced by geometry and solvation, see e.g. {Section 1.3 of the} supplementary information, and Refs.\cite{blondin-90,noodleman1996valence}).
In the HDE model\cite{Gillum1976,noodleman1992}, this electron is placed in either a single bonding-type or
antibonding-type orbital between the centers, splitting each Heisenberg level by a double-exchange contribution $B(S+1/2)$, and giving levels of the form
\begin{equation}
E(S) = JS(S+1) \pm B(S+1/2). \label{eq:hde}
\end{equation}
As discussed by Noodleman and Baerends\cite{Noodleman1984}, double-exchange stabilizes high-spin states.

\begin{figure}
\includegraphics[width=0.4\textwidth]{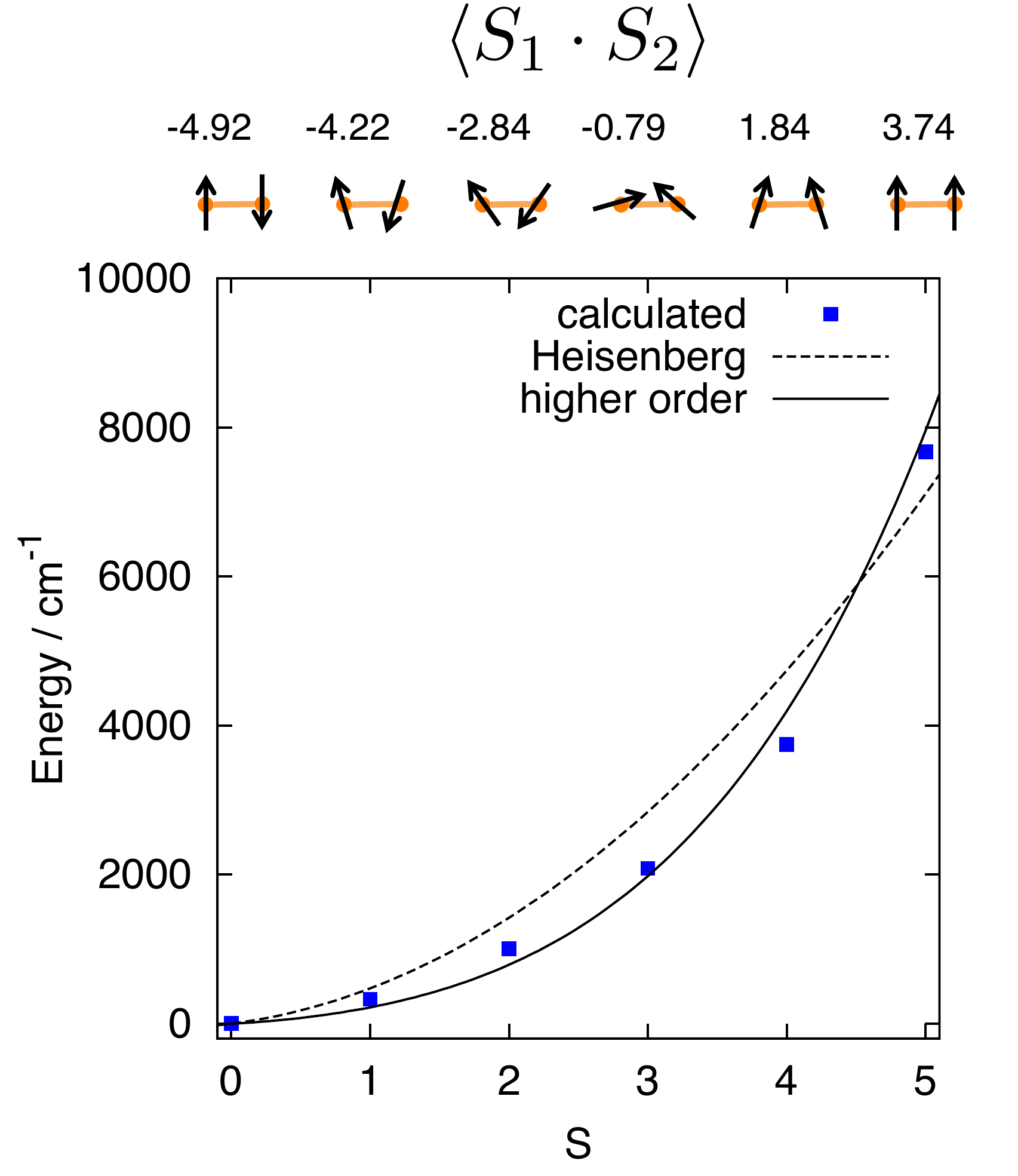}
\caption{{\it Ab-initio} levels of  [Fe$_2$S$_2$(SCH$_3$)$_4$]$^{2-}$
and corresponding model fits. (Dashed line) Fit to Heisenberg model.
(Solid line) Fit to Heisenberg plus quadratic couplings,  capturing charge fluctuation effects.
Top: Expectation value of the spin-correlation,  $\langle S_1\cdot S_2\rangle$, for increasing dimer spin, showing
 progression from antiferromagnetic to ferromagnetic ordering.\label{fig:fe2ox}}
\end{figure}

We now examine the accuracy of these existing pictures by comparing against the
electronic levels that can now be directly calculated using the {\it ab-initio} DMRG (details in {Sections 1.1 and 1.2 of the} supplementary information and
methods section). 
Figure \ref{fig:fe2ox} shows the computed levels of the oxidized { [2Fe-2S]$^{2-}$ } complex as compared to the predictions of the 
Heisenberg model, Eq.~(\ref{eq:heis}). The exact DMRG levels qualitatively form a spin ladder from $S$=0--5, in general
agreement with the Heisenberg model. As $S$ increases, the 
spin-spin correlation $\langle S_1\cdot S_2\rangle$ increases, transitioning from antiferromagnetic to ferromagnetic alignment.
Fitting to Eq.~(\ref{eq:heis}) yields $J\approx 236$ cm$^{-1}$, which
compares reasonably well to fits from magnetic susceptibility measurements on a similar synthetic dimer ($J\approx$ 148$\pm$16 cm$^{-1}$ \cite{Gillum1976}) and 
computed  BS-DFT estimates ($J\approx$ 310 cm$^{-1}$ \cite{noodleman1992}). 
However, our {\it ab-initio} levels also show significant deviations from the level structure assumed by the traditional Heisenberg model.
For example,  the Heisenberg model overestimates the lower spin state energies, while underestimating those of the higher spin states. Measuring the local spin on the Fe atoms
in our calculations we find $\langle S^2_1\rangle$ ranges from 5.47--5.74 (for the  $S$=0, $S$=5 states respectively), as compared with $\langle S_1^2\rangle$=8.75 for the pure Fe$^{\mathrm{III}}$ ($S$=5/2) ion assumed in the model. This deviation from a pure $S$=5/2 ion 
illustrates the Heisenberg model's limits, which does not allow for additional spin or charge configuration mixing. 
In fact, charge fluctuations are responsible for the important spin delocalization onto the S orbitals, as previously 
observed by Noodleman\cite{noodleman1992}. Guihery and coworkers have derived the form of the corrections to the Heisenberg model that arise from quantum charge fluctuations\cite{guihery}. In dimers, this yields a quadratic spin coupling $4J_Q (S_1\cdot S_2 )^2$; fitting to the {\it ab-initio} DMRG levels gives $J\approx 98$ cm$^{-1}$ 
and $J_Q\approx 6$ cm$^{-1}$.  As demonstrated in Figure \ref{fig:fe2ox}, the quadratic coupling greatly improves the agreement, showing the importance of these corrections.

\begin{figure}
\includegraphics[width=0.8\textwidth]{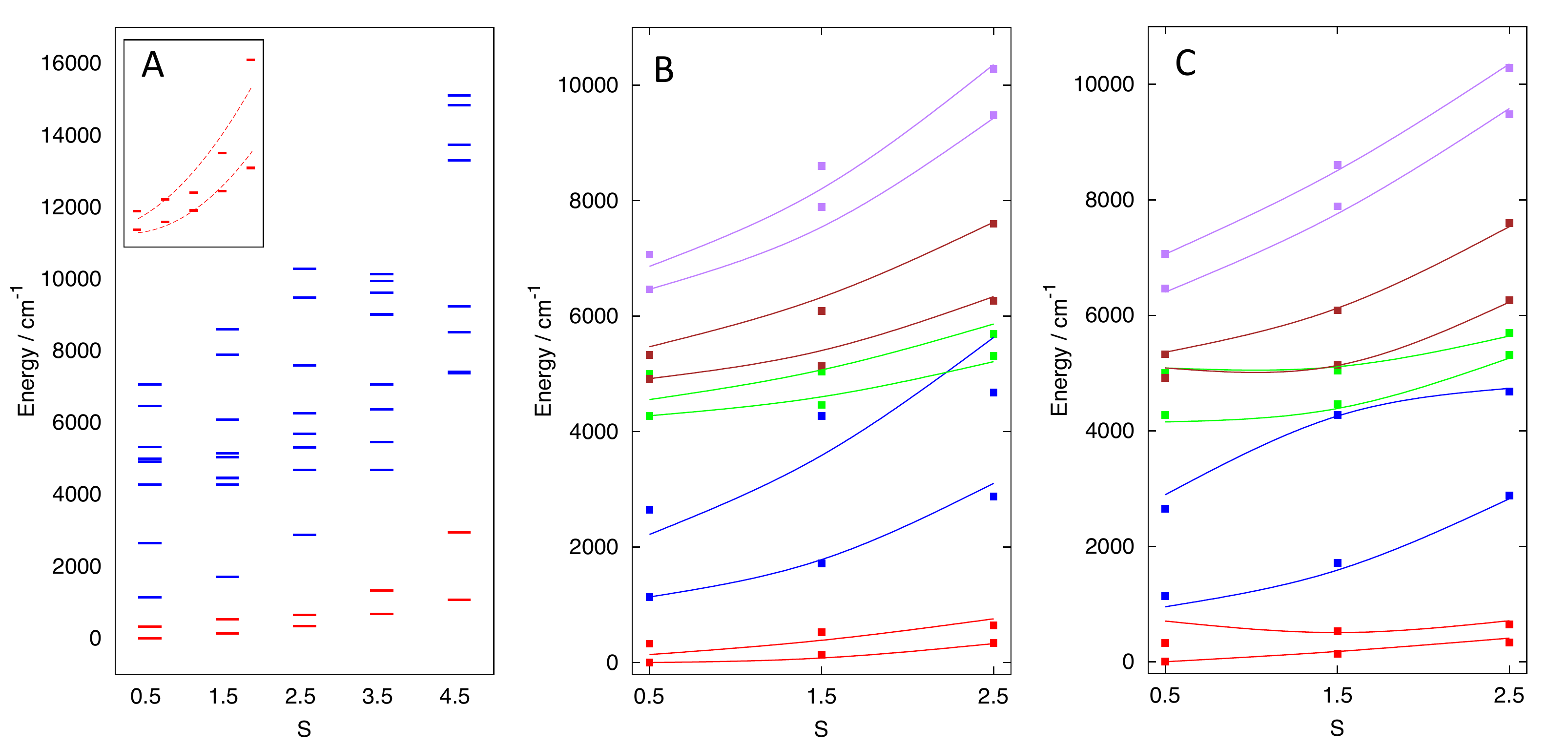}
\caption{{ {\it Ab-initio} levels of [Fe$_2$S$_2$(SCH$_3$)$_4$]$^{3-}$ and
corresponding model fits. The multi-orbital Anderson model proposed here produces an excellent fit to all the  levels, capturing the essential physics of the low energy states, while the 
single-pair and multi-pair HDE models do not.}  (A)  {\it Ab-initio} levels of [Fe$_2$S$_2$(SCH$_3$)$_4$]$^{3-}$ for each dimer spin $S$. (Inset: Fit of lowest two levels (red) to the HDE model).
The level separation does not increase monotonically as required in the HDE model, and 
the separation between the lowest two and higher levels is small (especially for  $S$=1/2, 3/2, 5/2)
indicating that the HDE model assumptions break down and multi-orbital double-exchange is important.
(B) Fits ($S$=1/2, 3/2, 5/2) to a multi-pair HDE model. {Pairs of corresponding bonding and anti-bonding states are represented by the same color.} 
(C) Fits ($S$=1/2, 3/2, 5/2) to a multi-orbital Anderson model. (B) and (C) use the same number of parameters, but (C)
is much better than (B), demonstrating that multi-orbital double-exchange is not a sum of single exchange processes, as required by the multi-pair HDE model.\label{fig:fe3spec}}
\end{figure}

\begin{figure}
\includegraphics[width=0.4\textwidth]{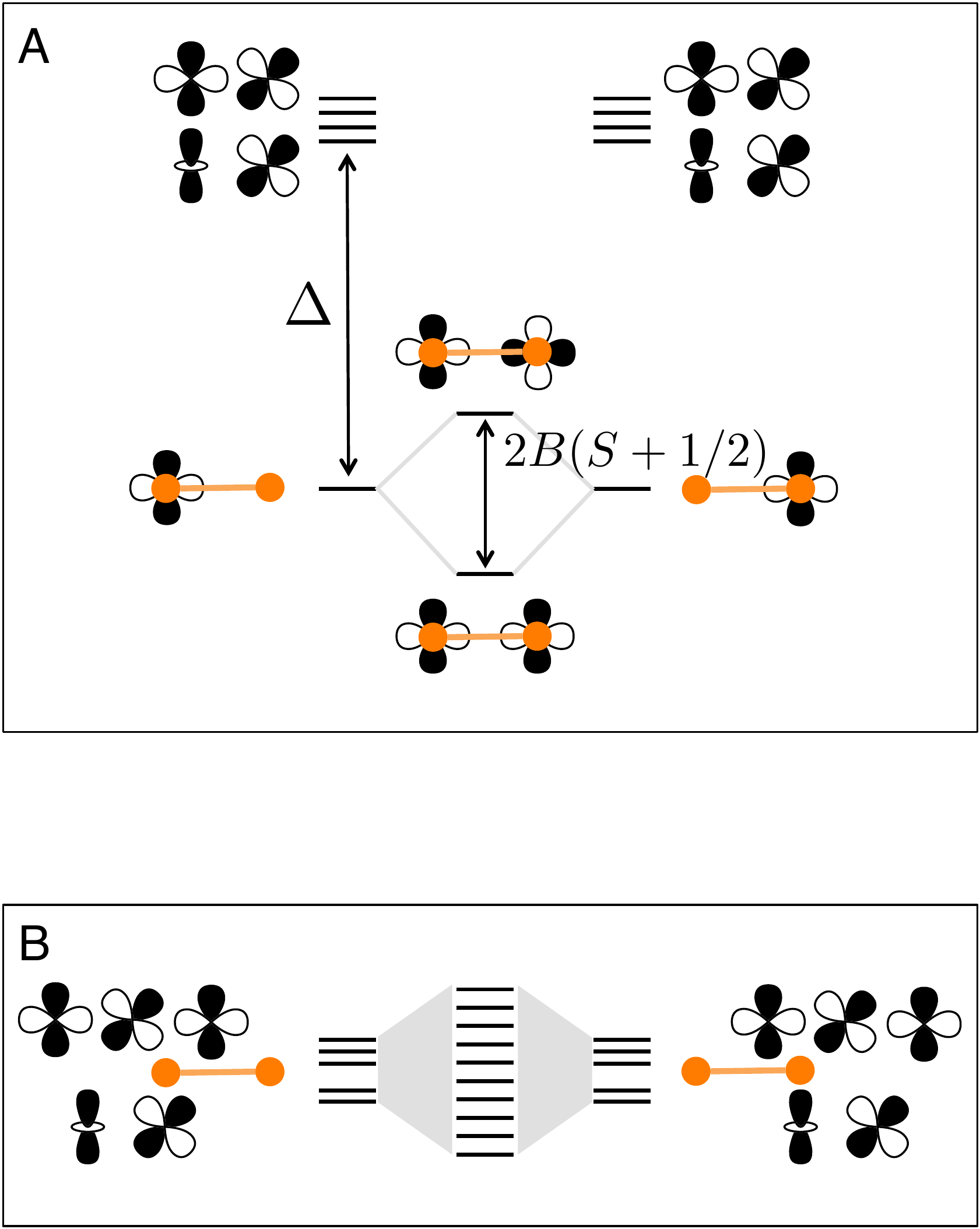}
\caption{{Origin of the dense manifold in the reduced dimer. Although the HDE model predicts two low-lying states in the reduced dimer, the observed dense manifold can result from a ligand field splitting that
 is energetically comparable to the hopping energy.}
(A) The HDE model assumes a single electron hops between a pair of d orbitals on the Fe ions. 
This is valid if $\Delta$ is sufficiently 
large that  other d orbitals are well separated. (B) In FeS systems, $\Delta$ is comparable
to the hopping energy, giving a whole manifold of  low-lying states at each dimer spin.\label{fig:fe3hopping}}
\end{figure}

We now turn to the reduced { [2Fe-2S]$^{3-}$ } dimer. Whereas the disagreement between the standard HDE model, and the directly computed {\it ab-initio} spectrum 
for the oxidized dimer, was primarily quantitative, for the reduced dimer the discrepancies are more severe. The HDE model predicts the splitting of the two lowest levels 
to increase with total spin, $2B(S+1/2)$. However,  in the {\it ab-initio} spectrum (Figure \ref{fig:fe3spec}, panel A, red-curves)  the splitting {\it decreases} with dimer spin from $S$=1/2--3/2. Clearly, this cannot be reproduced by {\it any} HDE model parameters, since the splitting is always proportional to $S$. 
(This remains after geometry relaxation, which introduces trapping (see {Section 1.3 of the} supplementary information)).

That the earlier model description breaks down for highly excited states is natural, but that it fails already for the lowest 
two states is surprising. As explained further below, this reflects the complicated nature of double-exchange in nearly orbitally degenerate 
metal ions. The HDE model assumes that the additional hopping 
electron is held in a single pair of prescribed bonding and antibonding orbitals bridging the
 the sulfur ligands. This is appropriate if  ligand-field splitting places other bonding and antibonding orbitals at 
higher energies (panel A of Figure \ref{fig:fe3hopping}), but is not in fact the 
 case for tetrahedrally coordinated Fe$^\mathrm{III}$ ions where both orbitals of $e$-parentage (in $T_d$ symmetry) are near-degenerate,
and weak ligand-field splitting by  $\pi$-donor ligands  leads to  higher-lying orbitals of $t_2$-parentage also being
accessible for double-exchange (panel B of Figure \ref{fig:fe3hopping}). 

The multi-orbital nature of Fe double-exchange is immediately seen in the level spectrum at each dimer spin $S$ (Figure \ref{fig:fe3spec}).
Especially for $S$=1/2, 3/2, 5/2, the separation between the first two states and the higher states is small,  rendering
the single pair HDE model completely invalid. Instead, the lowest 10 levels constitute a dense manifold of related states, where the hopping electron occupies any one of 10 available d orbitals 
on the Fe atoms. 

Multi-orbital double-exchange explains many of the earlier difficulties in determining consistent double-exchange parameters from experiment.
Simply put, a well-defined global $B$ does not exist. 
Fitting the two lowest curves to the HDE model yields $J\approx 67$ cm$^{-1}$, and 
$B\approx 63$ cm$^{-1}$, while
fitting the lowest and highest curves in the manifold, yields $J \approx 311$ cm$^{-1}$ and $B\approx 1052$ cm$^{-1}$. 
This wide range in $B$ values is consistent with earlier experimental fits, which also yield $B$ couplings varying by a factor of 50 or more \cite{Beinert-science}. 
(For a careful analysis of experimental variation in $B$, see Ref. \cite{noodleman1996valence}).

A naive way to extend the canonical HDE model to include multi-orbital exchange
is to assume that the HDE energy levels in Eq. (\ref{eq:hde}) simply generalize from 1 to 5 hopping pairs and
5 sets of $J_i$ and $B_i$, $i=1\ldots 5$ couplings, with ligand-field splittings $\Delta_i$, $i=1\ldots 4$.
 (Multi-pair HDE models have been considered for [4Fe-4S] clusters in the context
of $\sigma$ versus $\delta$ delocalization pathways \cite{noodleman1992,Noodleman1985}).
This gives levels of the form
\begin{equation}
  E_i(S)=\Delta_i + 2J_i S_1 \cdot S_2 \pm B_i(S_i+1/2) \label{eq:hdegeneralized}
\end{equation}
where $i$ labels the orbital pair associated with the hopping.
Fits to this multi-pair form are shown in Figure \ref{fig:fe3spec} (panel B). The multi-pair HDE model does capture the low-lying spectrum better, since it has more parameters, but
it still cannot reproduce the non-monotonic behaviour in pairs of energy levels, such as the decreasing gap between the first two levels.
This shows that multi-orbital exchange cannot be considered a simple pairwise process.

The correct qualitative picture for multi-orbital exchange requires a return to 
Anderson's original Hamiltonian for double-exchange.
For this system it takes the form (detailed derivation {in Section 1.4 of} the supplementary information)
 \begin{equation}
  H = \sum_i J_i s_{1i} \cdot s_{2i}  
  + \sum_{i\sigma} \left[\beta_i (c_{1i\sigma}^\dag c_{2i\sigma} + c_{2i\sigma}^\dag c_{1i\sigma}) + \Delta_i (c_{1i\sigma}^\dag c_{1i\sigma} + c_{2i\sigma}^\dag c_{2i\sigma})\right]
  \label{eq:multiorbitalh}
\end{equation}
($i$ labels  $d$ orbitals, $s_{1i}$, $s_{2i}$ are electron spins, 
and $c_{1i}^{(\dag)}$, $c_{2i}^{(\dag)}$ create and destroy electrons).
Note that this Anderson Hamiltonian has the {\it same} number of parameters
as a multi-pair HDE model (Eq. \ref{eq:hdegeneralized}), but the fit 
(Figure \ref{fig:fe3spec}, panel C) is much improved and now obtains
the correct non-monotonic features. Further, the fit is stable, near unique, and yields 
reasonable parameters:  ligand field splittings are approximately 4000-5000
cm$^{-1}$, consistent with spectroscopic estimates for tetrahedral Fe (see {Supplementary Table 10}).
Thus all qualitative features of our {\it ab-initio}  spectrum can be understood by treating
multi-orbital exchange in this more complete way.

In summary, our directly computed levels show that the low-lying spectrum of [2Fe-2S] dimers is much richer and denser than the simple pair-splitting
 long assumed within the canonical HDE picture; the level spectrum is generated by a complex multi-orbital exchange process, and this process 
cannot be viewed as a simple ``sum'' over
single orbital double-exchange pathways.

\subsection{[4Fe-4S] clusters.}

We now turn from FeS dimers to the more complicated [4Fe-4S] clusters. 
As a representative of Nature's cubanes we consider the [Fe$_4$S$_4$(SCH$_3$)$_4$)]$^{2-}$ cluster, derived from the synthetic cluster studied by
Holm and coworkers \cite{Averill1973}. The deduction of the cubane ground-state from experimental measurements
is an early triumph of inorganic spectroscopy \cite{Papa-mauss}. It is conventionally believed that
the ground-state consists of two coupled iron dimers, located on opposite faces of the cube. 
The dimers are thought of as mixed-valence Fe$^{2.5+}$, Fe$^{2.5+}$ pairs  coupled
to a high-spin $S$=9/2 state, which further recouple in the ground-state to form an overall singlet. 

\begin{figure}
\includegraphics[width=0.4\textwidth]{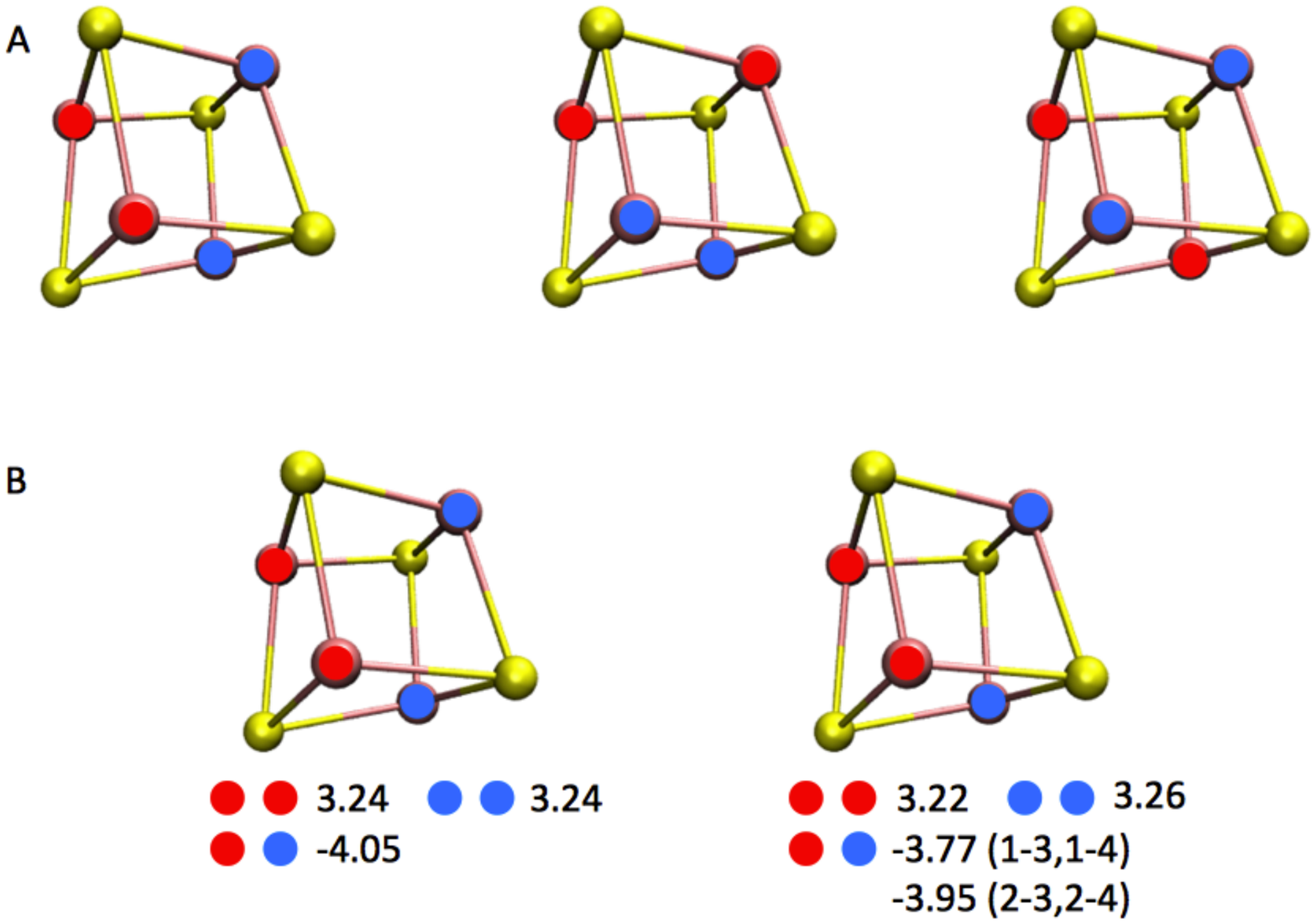}
\caption{{Spin pairings in the [4Fe-4S] cluster. Unlike in a perfectly cubic cluster, where 3 equivalent pairings of Fe's exist, the geometry-optimized cluster breaks this symmetry and favors one pairing over the other two. The dimer pairs, and their relative spin orientations, can be detected through the spin correlation functions.} (A) 3 equivalent pairings of Fe's into dimers in a perfect [4Fe-4S] cubane. Red-red: Ferromagnetic coupling, red-blue: Antiferromagnetic coupling.
(B) Spin correlation functions ($\langle S_i \cdot S_j\rangle$) for lowest singlet (left) and triplet (right) states of the [Fe$_4$S$_4$(SCH$_3$)$_4$]$^{2-}$ cluster. \label{fig:fe4pairs}}
\end{figure}

Since we can compute the electronic structure of the [4Fe-4S] cluster directly (see Methods) we can now test the long-standing hypothesis for the ground-state.
In our computed singlet ground-state, the spin density is zero everywhere (since our wavefunction is a true spin singlet), but the nature of the 
couplings can be established from spin correlation functions $\langle S_i \cdot S_j\rangle$ (Figure \ref{fig:fe4pairs}). These indicate
ferromagnetic coupling along the top and bottom faces with antiferromagnetic coupling between the faces, 
corresponding precisely to the experimental picture of two high-spin Fe$^{2.5+}$, Fe$^{2.5+}$ dimers recoupled into a singlet, thus confirming
the long-standing interpretation. We can further extend our calculations to the lowest triplet state:
 this lies 351 cm$^{-1}$ above the singlet state. From its spin correlation functions, we see that the triplet corresponds to a spin-canted state: the $S$=9/2 dimer spins are tilted relative to their ground-state orientation.

\begin{figure}
\includegraphics[width=0.6\textwidth]{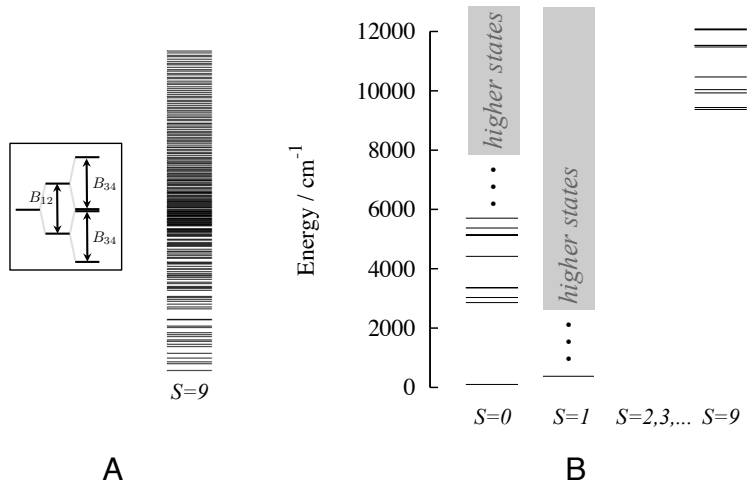}
\caption{{Energy levels of the [4Fe-4S] cluster. The low-lying spectrum of the [4Fe-4S] cluster is unusually dense and is starkly different from the predictions of the simple HDE model.} (A) Isolated 4 low-lying states predicted (for each $S$) from the HDE model, 
compared to the qualitative {\it ab-initio}  density of states (shown for $S$=9), (B) 
Lowest lying 10 singlet ($S$=0) and high spin ($S$=9) states), as well as ground-state triplet $S$=1 state,
showing the small energy scales and detailed structure of the spectrum.\label{fig:fe4spec}}
\end{figure}

What about the complete [4Fe-4S] spectrum? From the complexity of the [2Fe-2S] spectrum, we expect  this will be a formidable beast.
 %In the [2Fe-2S] dimer we already found 10 low-lying double-exchange states per dimer spin, while the standard HDE model predicted only two. In the [4Fe-4S] cluster we expect to see about the square of this number of low-lying double-exchange states {\it per cluster spin}, arising from the hopping electrons occupying any of the 20 Fe 3d orbitals, as compared to 4 low-lying levels per spin in the HDE model  (Figure \ref{fig:fe4spec}, left inset). 
While the complete spectrum is too expensive, we can calculate some of the lower-lying states. 
The 10 lowest singlet $S$=0 and high-spin $S$=9 states are shown in Figure~\ref{fig:fe4spec}, as well as the lowest triplet, and a (more qualitative) spectrum for the 150 lowest $S$=9 states.
In a perfect cubane we expect three degenerate singlet ground-states, from the three possible spin couplings in Figure \ref{fig:fe4pairs}.
However,  vibronic  coupling distorts the ground-state of the [4Fe-4S] cluster, opening a gap between the ground-
and higher-lying states. Neverthless, even in the distorted [4Fe-4S] cluster, the manifold of low-lying states remains accessible and dense on the 10-20~kcal/mol scale of 
biological FeS reorganization energies~\cite{Noodleman:2002:1367-5931:259}. A detailed analysis of the [4Fe-4S] excited states and their coupling to distortions is presented {in
Section 2.4 of the} supplementary information.
Further,  the  structure of the low-lying spectrum, for any cluster spin $S$, is very different to the isolated 4 low-lying levels predicted by canonical HDE models
for each $S$. This is expected as the  model fails already
 in the [2Fe-2S] dimers, due to  neglect of multi-orbital double exchange discussed above. 
Appropriate model Hamiltonians for the [4Fe-4S] clusters are analysed 
in detail in {Section 2.4 of the} supplementary information.

%%   (compared to the barrier height about 20 kcal/mole for C-H activation reaction performed by single-Fe porphyrins\cite{Shaik2005}) and has a very different structure than the 4 low-lying levels of the canonical
%% HDE model. }
%%  are calculated and plotted in Figure\ref{fig:fe4spec}. The density of states in the [4Fe-4S] cluster is less than that of the [2Fe-2S] cluster because of substantial vibronic coupling which stabilizes the ground state of our distorted [4Fe-4S] cluster relative to the higher lying states.  %While computing the complete singlet spectrum is too expensive, it is feasible for the high-spin $S$=9 states.
 
The high density of states (and their sensitivity to geometry via the vibronic coupling) provides an intriguing hypothesis
behind the unusual chemical flexibility and ubiquity of the [4Fe-4S] clusters. Conventionally, molecular reactivity  proceeds via
 well-defined  potential energy surfaces. However, in [4Fe-4S] clusters, a large number of states of the same (and different) spin are
energetically accessible during reorganization dynamics. This suggests that [4Fe-4S] clusters can non-adiabatically switch between many different frontier electronic states in a reaction, 
in essence a generalization of two-state reactivity already postulated in single-Fe porphyrins\cite{Shaik2005} but without slow spin-forbidden state crossings. 
%% In the case of electron transfer this is equivalent to a low electronic re-organization energy. 
Coupling between the spectrum and geometry allows for ``fine-tuning'' of reactivity by the environment.
[4Fe-4S] clusters thus appear to provide a conceptual bridge between molecular and surface catalytic reactivity. In the latter case, non-adiabatic processes on many potential energy surface 
are common, and greatly modify the timescales of electron and energy transfer\cite{Wodtke2004}.

\section{Conclusions}
To summarize,  using simplifying entanglement structure of physical many-particle quantum states, we have computed
 the individual ground and excited state energy levels of [2Fe-2S] and [4Fe-4S] clusters without model assumptions. Direct access to
these energy levels has not previously been possible either through experiment or theory. Our calculations have allowed us to critically examine the 
 validity of the consensus phenomenological models that have so far been the only way to
 understand FeS chemistry. In both the [2Fe-2S] dimer and the [4Fe-4S] cluster, we find that earlier
understanding based on the canonical Heisenberg-Double-Exchange model underestimates the number of low-lying states by 1-2 orders of magnitude. 
These low-lying states arise from multi-orbital double-exchange processes. 
The new level spectrum  we reveal has important implications for reactivity, as the density 
and accessibility of the low-lying states argues for the importance of
multiple electronic states and non-adiabaticity in reactions.
The theoretical techniques described here are potentially applicable to biological systems of even greater complexity, including the M- and P-clusters of 
nitrogenase \cite{Frank-Science,Spatzal18112011}. More broadly, our work demonstrates new possibilities of realising spectroscopy in complex systems, by directly
computing entangled electronic structure from many-particle quantum mechanics, without the need for a {\it priori} model assumptions.

\newpage

\section{Methods}
All our density matrix renormalization group (DMRG) calculations employed the \textsc{Block} code. Density functional calculations to obtain the initial geometries were carried out using the \textsc{ORCA} package.
For DMRG calculations on the [2Fe-2S] clusters, we employed
a full valence complete active space (all Fe 3d and S 3p orbitals), and further include the Fe 4s and 4d 
shells to account for additional dynamic correlation effects. In the oxidized cluster, this corresponds to a 30 electron, 32 orbital active space. Even though the formal Hilbert space dimension is greater than 10$^{17}$, our calculations are enabled by  the presence of some
special entanglement structure in the states.
In the [2Fe-2S] clusters, we estimate that the electronic relative energies are converged to better than 0.1 kcal/mol ($\approx$ 35 cm$^{-1}$)
of the exact active space results. For the [4Fe-4S] clusters we employed an active space with all Fe 3d and S 3p orbitals.
These calculations were more expensive than the dimer calculations, thus the DMRG energy differences between the singlet and triplet states
are converged to only about 0.5-1 kcal/mol. 
These estimated errors refer to the errors from the corresponding complete active space result. Dynamical correlation may lead to further small  changes, as discussed in {the Section 1.2 of} supplementary information.
The higher  singlet and high-spin states in Figure~\ref{fig:fe4spec} were 
calculated to lower accuracy than the ground singlet and triplet states, but are qualitatively correct.
The multiplet states are computed  explicitly without assuming any  mutual inter-relationship, in contrast to BS-DFT techniques, which only obtain the high-spin state and a weighted average
 of the multiplet energies, relying on model assumptions to deduce the individual levels \cite{Neese2009}. 
All our computed wavefunctions exactly preserve spin ($S^2$) symmetry, due to the use of an SU(2) invariant DMRG code developed in our lab\cite{Sharma2012}. A full discussion of all the active spaces, geometry, and examination of the DMRG convergence,  is presented in the {Sections 1.1, 1.2, 2.1 and 2.2 of the }supplementary information.
For the spin correlation functions, $S_1$ and $S_2$ are defined from local Fe spin operators, and $\langle S_1\cdot S_2\rangle$ is the {\it ab-initio} expectation value.
These operators are defined  in {Equations 1 and 2 of} the supplementary information.

\section{Acknowledgements} Work performed by S.S. and G.K.C. was supported by the US National Science Foundation CHE-1265277, using software developed with
the support of OCI-1265278. F.N. and K.S. gratefully acknowledge financial support of this work by the Max Planck Society, the University of Bonn and the SFB 813 "Chemistry at Spin Centers".
{
\section{Author Contributions}
S.S. performed the DMRG calculations, analyzed the results and contributed to the writing of the manuscript. F.N. contributed to the writing of the manuscript. K.S. performed the geometry optimization for the [4Fe-4S] model cluster. G.K.C. wrote the manuscript and contributed to the calculations and analysis of the results. All authors discussed the results and commented on the manuscript.}
%\bibliography{Fe-S}

\begin{thebibliography}{10}
\expandafter\ifx\csname url\endcsname\relax
  \def\url#1{\texttt{#1}}\fi
\expandafter\ifx\csname urlprefix\endcsname\relax\def\urlprefix{URL }\fi
\providecommand{\bibinfo}[2]{#2}
\providecommand{\eprint}[2][]{\url{#2}}

\bibitem{Beinert-science}
\bibinfo{author}{Beinert, H.}, \bibinfo{author}{Holm, R.~H.} \&
  \bibinfo{author}{M\"{u}nck, E.}
\newblock \bibinfo{title}{{Iron-sulfur clusters: nature's modular, multipurpose
  structures}}.
\newblock \emph{\bibinfo{journal}{Science}} \textbf{\bibinfo{volume}{277}},
  \bibinfo{pages}{653--659} (\bibinfo{year}{1997}).

\bibitem{Johnson2005}
\bibinfo{author}{Johnson, D.~C.}, \bibinfo{author}{Dean, D.~R.},
  \bibinfo{author}{Smith, A.~D.} \& \bibinfo{author}{Johnson, M.~K.}
\newblock \bibinfo{title}{{Structure, function, and formation of biological
  iron-sulfur clusters}}.
\newblock \emph{\bibinfo{journal}{Annual Review of Biochemistry}}
  \textbf{\bibinfo{volume}{74}}, \bibinfo{pages}{247--281}
  (\bibinfo{year}{2005}).

\bibitem{Noodleman1995}
\bibinfo{author}{Noodleman, L.}, \bibinfo{author}{Peng, C.~Y.},
  \bibinfo{author}{Case, D.~A.} \& \bibinfo{author}{Mouesca, J.~M.}
\newblock \bibinfo{title}{{Orbital interactions, electron delocalization and
  spin coupling in iron-sulfur clusters}}.
\newblock \emph{\bibinfo{journal}{Coordination Chemistry Reviews}}
  \textbf{\bibinfo{volume}{144}}, \bibinfo{pages}{199--244}
  (\bibinfo{year}{1995}).

\bibitem{zener}
\bibinfo{author}{Zener, C.}
\newblock \bibinfo{title}{{Interaction between the d-shells in the transition
  metals. II. Ferromagnetic compounds of manganese with perovskite structure}}.
\newblock \emph{\bibinfo{journal}{Phys. Rev.}} \textbf{\bibinfo{volume}{82}},
  \bibinfo{pages}{403--405} (\bibinfo{year}{1951}).

\bibitem{anderson-DE}
\bibinfo{author}{Anderson, P.~W.} \& \bibinfo{author}{Hasegawa, H.}
\newblock \bibinfo{title}{{Considerations on double exchange}}.
\newblock \emph{\bibinfo{journal}{Physical Review}}
  \textbf{\bibinfo{volume}{100}}, \bibinfo{pages}{675--681}
  (\bibinfo{year}{1955}).

\bibitem{girerd}
\bibinfo{author}{Girerd, J.-J.}
\newblock \bibinfo{title}{{Electron transfer between magnetic ions in mixed
  valence binuclear systems}}.
\newblock \emph{\bibinfo{journal}{The Journal of Chemical Physics}}
  \textbf{\bibinfo{volume}{79}}, \bibinfo{pages}{1766--1775}
  (\bibinfo{year}{1983}).

\bibitem{Noodleman1984}
\bibinfo{author}{Noodleman, L.} \& \bibinfo{author}{Baerends, E.~J.}
\newblock \bibinfo{title}{{Electronic structure, magnetic properties, ESR, and
  optical spectra for 2-iron ferredoxin models by LCAO-X$\alpha$ valence bond
  theory}}.
\newblock \emph{\bibinfo{journal}{Journal of the American Chemical Society}}
  \textbf{\bibinfo{volume}{106}}, \bibinfo{pages}{2316--2327}
  (\bibinfo{year}{1984}).

\bibitem{noodleman1986ligand}
\bibinfo{author}{Noodleman, L.} \& \bibinfo{author}{Davidson, E.~R.}
\newblock \bibinfo{title}{Ligand spin polarization and antiferromagnetic
  coupling in transition metal dimers}.
\newblock \emph{\bibinfo{journal}{Chemical physics}}
  \textbf{\bibinfo{volume}{109}}, \bibinfo{pages}{131--143}
  (\bibinfo{year}{1986}).

\bibitem{papa-fe3}
\bibinfo{author}{Papaefthymiou, V.}, \bibinfo{author}{Girerd, J.~J.},
  \bibinfo{author}{Moura, I.}, \bibinfo{author}{Moura, J. J.~G.} \&
  \bibinfo{author}{Muenck, E.}
\newblock \bibinfo{title}{{Moessbauer study of D. gigas ferredoxin II and
  spin-coupling model for Fe$_3$S$_4$ cluster with valence delocalization}}.
\newblock \emph{\bibinfo{journal}{Journal of the American Chemical Society}}
  \textbf{\bibinfo{volume}{109}}, \bibinfo{pages}{4703--4710}
  (\bibinfo{year}{1987}).

\bibitem{blondin-90}
\bibinfo{author}{Blondin, G.} \& \bibinfo{author}{Girerd, J.~J.}
\newblock \bibinfo{title}{{Interplay of electron exchange and electron transfer
  in metal polynuclear complexes in proteins or chemical models}}.
\newblock \emph{\bibinfo{journal}{Chemical Reviews}}
  \textbf{\bibinfo{volume}{90}}, \bibinfo{pages}{1359--1376}
  (\bibinfo{year}{1990}).

\bibitem{Borshch1984}
\bibinfo{author}{Borshch, S.~A.}, \bibinfo{author}{Kotov, I.~N.} \&
  \bibinfo{author}{Bersuker, I.~B.}
\newblock \bibinfo{title}{{A vibronic model for exchange-coupled mixed-valence
  dimers}}.
\newblock \emph{\bibinfo{journal}{Chemical Physics Letters}}
  \textbf{\bibinfo{volume}{111}}, \bibinfo{pages}{264--270}
  (\bibinfo{year}{1984}).

\bibitem{guihery}
\bibinfo{author}{Labeguerie, P.} \emph{et~al.}
\newblock \bibinfo{title}{{Is it possible to determine rigorous magnetic
  Hamiltonians in spin s = 1 systems from density functional theory
  calculations?}}
\newblock \emph{\bibinfo{journal}{The Journal of Chemical Physics}}
  \textbf{\bibinfo{volume}{129}}, \bibinfo{pages}{154110}
  (\bibinfo{year}{2008}).

\bibitem{Gibson}
\bibinfo{author}{Gibson, J.~F.}, \bibinfo{author}{Hall, D.~O.},
  \bibinfo{author}{Thornley, J.~H.} \& \bibinfo{author}{Whatley, F.~R.}
\newblock \bibinfo{title}{{The iron complex in spinach ferredoxin}}.
\newblock \emph{\bibinfo{journal}{Proceedings of the National Academy of
  Sciences}} \textbf{\bibinfo{volume}{56}}, \bibinfo{pages}{987--990}
  (\bibinfo{year}{1966}).

\bibitem{Sands}
\bibinfo{author}{Brintzinger, H.}, \bibinfo{author}{Palmer, G.} \&
  \bibinfo{author}{Sands, R.~H.}
\newblock \bibinfo{title}{{On the ligand field of iron in ferredoxin from
  spinach chloroplasts and related nonheme iron enzymes}}.
\newblock \emph{\bibinfo{journal}{Proceedings of the National Academy of
  Sciences}} \textbf{\bibinfo{volume}{55}}, \bibinfo{pages}{397--404}
  (\bibinfo{year}{1966}).

\bibitem{Rius89}
\bibinfo{author}{Rius, G.} \& \bibinfo{author}{Lamotte, B.}
\newblock \bibinfo{title}{{Single-crystal ENDOR study of an $^{57}$Fe-enriched
  iron-sulfur [Fe$_4$S$_4$]$^{3+}$ cluster}}.
\newblock \emph{\bibinfo{journal}{Journal of the American Chemical Society}}
  \textbf{\bibinfo{volume}{111}}, \bibinfo{pages}{2464--2469}
  (\bibinfo{year}{1989}).

\bibitem{Mouesca1991}
\bibinfo{author}{Mouesca, J.~M.}, \bibinfo{author}{Lamotte, B.} \&
  \bibinfo{author}{Rius, G.}
\newblock \bibinfo{title}{{Comparison between spin population distributions in
  two different [Fe$_4$S$_4$]$^{3+}$ clusters by proton ENDOR in single
  crystals of a synthetic model compound.}}
\newblock \emph{\bibinfo{journal}{Journal of Inorganic Biochemistry}}
  \textbf{\bibinfo{volume}{43}}, \bibinfo{pages}{251} (\bibinfo{year}{1991}).

\bibitem{Bertini1991}
\bibinfo{author}{Bertini, I.}, \bibinfo{author}{Briganti, F.},
  \bibinfo{author}{Luchinat, C.}, \bibinfo{author}{Scozzafava, A.} \&
  \bibinfo{author}{Sola, M.}
\newblock \bibinfo{title}{{Proton NMR spectroscopy and the electronic structure
  of the high potential iron-sulfur protein from Chromatium vinosum}}.
\newblock \emph{\bibinfo{journal}{Journal of the American Chemical Society}}
  \textbf{\bibinfo{volume}{113}}, \bibinfo{pages}{1237--1245}
  (\bibinfo{year}{1991}).

\bibitem{Banci1991}
\bibinfo{author}{Banci, L.} \emph{et~al.}
\newblock \bibinfo{title}{{Proton NMR spectra of oxidized high-potential
  iron-sulfur protein (HiPIP) from Rhodocyclus gelatinosus. A model for
  oxidized HiPIPs}}.
\newblock \emph{\bibinfo{journal}{Inorganic Chemistry}}
  \textbf{\bibinfo{volume}{30}}, \bibinfo{pages}{4517--4524}
  (\bibinfo{year}{1991}).

\bibitem{Papa-mauss}
\bibinfo{author}{Papaefthymiou, V.}, \bibinfo{author}{Millar, M.~M.} \&
  \bibinfo{author}{Muenck, E.}
\newblock \bibinfo{title}{{Moessbauer and EPR studies of a synthetic analog for
  the iron-sulfur Fe$_4$S$_4$ core of oxidized and reduced high-potential iron
  proteins}}.
\newblock \emph{\bibinfo{journal}{Inorganic Chemistry}}
  \textbf{\bibinfo{volume}{25}}, \bibinfo{pages}{3010--3014}
  (\bibinfo{year}{1986}).

\bibitem{Kappl2006}
\bibinfo{author}{Kappl, R.}, \bibinfo{author}{Ebelsh\"{a}user, M.},
  \bibinfo{author}{Hannemann, F.}, \bibinfo{author}{Bernhardt, R.} \&
  \bibinfo{author}{H\"{u}ttermann, J.}
\newblock \bibinfo{title}{{Probing electronic and structural properties of the
  reduced [2Fe--2S] cluster by orientation-selective $^1$H ENDOR spectroscopy:
  Adrenodoxin versus Rieske iron-sulfur protein}}.
\newblock \emph{\bibinfo{journal}{Applied Magnetic Resonance}}
  \textbf{\bibinfo{volume}{30}}, \bibinfo{pages}{427--459}
  (\bibinfo{year}{2006}).

\bibitem{Maurice2009}
\bibinfo{author}{Maurice, R.}, \bibinfo{author}{GuiheÃÅry, N.},
  \bibinfo{author}{Bastardis, R.} \& \bibinfo{author}{de~Graaf, C.}
\newblock \bibinfo{title}{{Rigorous extraction of the anisotropic multispin
  hamiltonian in bimetallic complexes from the exact electronic Hamiltonian}}.
\newblock \emph{\bibinfo{journal}{Journal of Chemical Theory and Computation}}
  \textbf{\bibinfo{volume}{6}}, \bibinfo{pages}{55--65} (\bibinfo{year}{2009}).

\bibitem{Noodleman1985}
\bibinfo{author}{Noodleman, L.}, \bibinfo{author}{Norman, J.~G.},
  \bibinfo{author}{Osborne, J.~H.}, \bibinfo{author}{Aizman, A.} \&
  \bibinfo{author}{Case, D.~A.}
\newblock \bibinfo{title}{{Models for ferredoxins: electronic structures of
  iron-sulfur clusters with one, two, and four iron atoms}}.
\newblock \emph{\bibinfo{journal}{Journal of the American Chemical Society}}
  \textbf{\bibinfo{volume}{107}}, \bibinfo{pages}{3418--3426}
  (\bibinfo{year}{1985}).

\bibitem{mouesca1994density}
\bibinfo{author}{Mouesca, J.-M.}, \bibinfo{author}{Chen, J.~L.},
  \bibinfo{author}{Noodleman, L.}, \bibinfo{author}{Bashford, D.} \&
  \bibinfo{author}{Case, D.~A.}
\newblock \bibinfo{title}{Density functional/poisson-boltzmann calculations of
  redox potentials for iron-sulfur clusters}.
\newblock \emph{\bibinfo{journal}{Journal of the American Chemical Society}}
  \textbf{\bibinfo{volume}{116}}, \bibinfo{pages}{11898--11914}
  (\bibinfo{year}{1994}).

\bibitem{shoji2007theory}
\bibinfo{author}{Shoji, M.} \emph{et~al.}
\newblock \bibinfo{title}{{Theory of chemical bonds in metalloenzymes III: Full
  geometry optimization and vibration analysis of ferredoxin-type [2Fe--2S]
  cluster}}.
\newblock \emph{\bibinfo{journal}{International Journal of Quantum Chemistry}}
  \textbf{\bibinfo{volume}{107}}, \bibinfo{pages}{116--133}
  (\bibinfo{year}{2007}).

\bibitem{noodleman81}
\bibinfo{author}{Noodleman, L.}
\newblock \bibinfo{title}{{Valence bond description of antiferromagnetic
  coupling in transition metal dimers}}.
\newblock \emph{\bibinfo{journal}{The Journal of Chemical Physics}}
  \textbf{\bibinfo{volume}{74}}, \bibinfo{pages}{5737--5743}
  (\bibinfo{year}{1981}).

\bibitem{Yamaguchi1989210}
\bibinfo{author}{Yamaguchi, K.}, \bibinfo{author}{Fueno, T.},
  \bibinfo{author}{Ueyama, N.}, \bibinfo{author}{Akira, N.} \&
  \bibinfo{author}{Masaaki, O.}
\newblock \bibinfo{title}{Antiferromagnetic spin couplings between iron ions in
  iron—sulfur clusters. a localized picture by the spin vector model}.
\newblock \emph{\bibinfo{journal}{Chemical Physics Letters}}
  \textbf{\bibinfo{volume}{164}}, \bibinfo{pages}{210 -- 216}
  (\bibinfo{year}{1989}).

\bibitem{Yamaguchi199056}
\bibinfo{author}{Yamaguchi, K.}, \bibinfo{author}{Fueno, T.},
  \bibinfo{author}{Ozaki, M.}, \bibinfo{author}{Ueyama, N.} \&
  \bibinfo{author}{Nakamura, A.}
\newblock \bibinfo{title}{A general spin-orbital (gso) description of
  antiferromagnetic spin couplings between four irons in iron-sulfur clusters}.
\newblock \emph{\bibinfo{journal}{Chemical Physics Letters}}
  \textbf{\bibinfo{volume}{168}}, \bibinfo{pages}{56 -- 62}
  (\bibinfo{year}{1990}).

\bibitem{Neese2009}
\bibinfo{author}{Neese, F.}
\newblock \bibinfo{title}{{Prediction of molecular properties and molecular
  spectroscopy with density functional theory: From fundamental theory to
  exchange-coupling}}.
\newblock \emph{\bibinfo{journal}{Coordination Chemistry Reviews}}
  \textbf{\bibinfo{volume}{253}}, \bibinfo{pages}{526--563}
  (\bibinfo{year}{2009}).

\bibitem{Miralles1992}
\bibinfo{author}{Miralles, J.}, \bibinfo{author}{Daudey, J.-P.} \&
  \bibinfo{author}{Caballol, R.}
\newblock \bibinfo{title}{{Variational calculation of small energy differences.
  The singlet-triplet gap in [Cu$_2$Cl$_6$]$_2$}}.
\newblock \emph{\bibinfo{journal}{Chemical Physics Letters}}
  \textbf{\bibinfo{volume}{198}}, \bibinfo{pages}{555--562}
  (\bibinfo{year}{1992}).

\bibitem{Miralles1993}
\bibinfo{author}{Miralles, J.}, \bibinfo{author}{Castell, O.},
  \bibinfo{author}{Caballol, R.} \& \bibinfo{author}{Malrieu, J.-P.}
\newblock \bibinfo{title}{{Specific CI calculation of energy differences:
  Transition energies and bond energies}}.
\newblock \emph{\bibinfo{journal}{Chemical Physics}}
  \textbf{\bibinfo{volume}{172}}, \bibinfo{pages}{33--43}
  (\bibinfo{year}{1993}).

\bibitem{Castell1999}
\bibinfo{author}{Castell, O.} \& \bibinfo{author}{Caballol, R.}
\newblock \bibinfo{title}{{Ab-initio configuration interaction calculation of
  the exchange coupling constant in hydroxo doubly bridged Cr(III) dimers}}.
\newblock \emph{\bibinfo{journal}{Inorganic Chemistry}}
  \textbf{\bibinfo{volume}{38}}, \bibinfo{pages}{668--673}
  (\bibinfo{year}{1999}).

\bibitem{Cabrero2000}
\bibinfo{author}{Cabrero, J.}, \bibinfo{author}{{Ben Amor}, N.},
  \bibinfo{author}{de~Graaf, C.}, \bibinfo{author}{Illas, F.} \&
  \bibinfo{author}{Caballol, R.}
\newblock \bibinfo{title}{{Ab-initio study of the exchange coupling in
  oxalato-bridged Cu(II) dinuclear complexes}}.
\newblock \emph{\bibinfo{journal}{The Journal of Physical Chemistry A}}
  \textbf{\bibinfo{volume}{104}}, \bibinfo{pages}{9983--9989}
  (\bibinfo{year}{2000}).

\bibitem{hubner}
\bibinfo{author}{H\"{u}bner, O.} \& \bibinfo{author}{Sauer, J.}
\newblock \bibinfo{title}{The electronic states of fe2s2$^{-/0/+/2+}$}.
\newblock \emph{\bibinfo{journal}{The Journal of Chemical Physics}}
  \textbf{\bibinfo{volume}{116}}, \bibinfo{pages}{617--628}
  (\bibinfo{year}{2002}).

\bibitem{hastings}
\bibinfo{author}{Hastings, M.~B.}
\newblock \bibinfo{title}{{Entropy and entanglement in quantum ground states}}.
\newblock \emph{\bibinfo{journal}{Phys. Rev. B}} \textbf{\bibinfo{volume}{76}},
  \bibinfo{pages}{35114} (\bibinfo{year}{2007}).

\bibitem{Verstraete2008}
\bibinfo{author}{Verstraete, F.}, \bibinfo{author}{Murg, V.} \&
  \bibinfo{author}{Cirac, J.~I.}
\newblock \bibinfo{title}{Matrix product states, projected entangled pair
  states, and variational renormalization group methods for quantum spin
  systems}.
\newblock \emph{\bibinfo{journal}{Adv. Phys.}} \textbf{\bibinfo{volume}{57}},
  \bibinfo{pages}{143} (\bibinfo{year}{2008}).

\bibitem{White1992}
\bibinfo{author}{White, S.~R.}
\newblock \bibinfo{title}{Density matrix formulation for quantum
  renormalization groups}.
\newblock \emph{\bibinfo{journal}{Phys. Rev. Lett.}}
  \textbf{\bibinfo{volume}{69}}, \bibinfo{pages}{2863} (\bibinfo{year}{1992}).

\bibitem{White1999}
\bibinfo{author}{White, S.~R.} \& \bibinfo{author}{Martin, R.~L.}
\newblock \bibinfo{title}{Ab initio quantum chemistry using the density matrix
  renormalization group}.
\newblock \emph{\bibinfo{journal}{J. Chem. Phys.}}
  \textbf{\bibinfo{volume}{110}}, \bibinfo{pages}{4127} (\bibinfo{year}{1999}).

\bibitem{chan2011}
\bibinfo{author}{Chan, G. K.-L.} \& \bibinfo{author}{Sharma, S.}
\newblock \bibinfo{title}{{The Density Matrix Renormalization Group in Quantum
  Chemistry}}.
\newblock \emph{\bibinfo{journal}{Ann. Rev. Phys. Chem.}}
  \textbf{\bibinfo{volume}{62}}, \bibinfo{pages}{465} (\bibinfo{year}{2011}).

\bibitem{Kurashige2013}
\bibinfo{author}{Kurashige, Y.}, \bibinfo{author}{Chan, G. K.-L.} \&
  \bibinfo{author}{Yanai, T.}
\newblock \bibinfo{title}{{Entangled quantum electronic wavefunctions of the
  Mn$_4$CaO$_5$ cluster in photosystem II}}.
\newblock \emph{\bibinfo{journal}{Nat Chem}} \textbf{\bibinfo{volume}{5}},
  \bibinfo{pages}{660--666} (\bibinfo{year}{2013}).

\bibitem{Sharma2012}
\bibinfo{author}{Sharma, S.} \& \bibinfo{author}{Chan, G. K.-L.}
\newblock \bibinfo{title}{{Spin-adapted density matrix renormalization group
  algorithms for quantum chemistry}}.
\newblock \emph{\bibinfo{journal}{The Journal of Chemical Physics}}
  \textbf{\bibinfo{volume}{136}}, \bibinfo{pages}{124121}
  (\bibinfo{year}{2012}).

\bibitem{Chan2002}
\bibinfo{author}{Chan, G. K.~L.} \& \bibinfo{author}{Head-Gordon, M.}
\newblock \bibinfo{title}{Highly correlated calculations with a polynomial cost
  algorithm: A study of the density matrix renormalization group}.
\newblock \emph{\bibinfo{journal}{J. Chem. Phys.}}
  \textbf{\bibinfo{volume}{116}}, \bibinfo{pages}{4462} (\bibinfo{year}{2002}).

\bibitem{Zgid2008spin}
\bibinfo{author}{Zgid, D.} \& \bibinfo{author}{Nooijen, M.}
\newblock \bibinfo{title}{On the spin and symmetry adaptation of the density
  matrix renormalization group method}.
\newblock \emph{\bibinfo{journal}{J. Chem. Phys.}}
  \textbf{\bibinfo{volume}{128}}, \bibinfo{pages}{014107}
  (\bibinfo{year}{2008}).

\bibitem{Moritz2005orb}
\bibinfo{author}{Moritz, G.}, \bibinfo{author}{Hess, B.~A.} \&
  \bibinfo{author}{Reiher, M.}
\newblock \bibinfo{title}{Convergence behavior of the density-matrix
  renormalization group algorithm for optimized orbital orderings}.
\newblock \emph{\bibinfo{journal}{J. Chem. Phys.}}
  \textbf{\bibinfo{volume}{122}}, \bibinfo{pages}{024107}
  (\bibinfo{year}{2005}).

\bibitem{kurashige}
\bibinfo{author}{Kurashige, Y.} \& \bibinfo{author}{Yanai, T.}
\newblock \bibinfo{title}{{High-performance ab initio density matrix
  renormalization group method: applicability to large-scale multireference
  problems for metal compounds}}.
\newblock \emph{\bibinfo{journal}{J. Chem. Phys.}}
  \textbf{\bibinfo{volume}{130}}, \bibinfo{pages}{234114}
  (\bibinfo{year}{2009}).

\bibitem{Legeza2003lif}
\bibinfo{author}{Legeza, {\"O}.}, \bibinfo{author}{R{\"o}der, J.} \&
  \bibinfo{author}{Hess, B.~A.}
\newblock \bibinfo{title}{{QC-DMRG} study of the ionic-neutral curve crossing
  of {LiF}}.
\newblock \emph{\bibinfo{journal}{Mol. Phys.}} \textbf{\bibinfo{volume}{101}},
  \bibinfo{pages}{2019} (\bibinfo{year}{2003}).

\bibitem{marti}
\bibinfo{author}{Marti, K.~H.}, \bibinfo{author}{Ondik, I.~M.},
  \bibinfo{author}{Mortise, G.} \& \bibinfo{author}{Reiher, M.}
\newblock \bibinfo{title}{Density matrix renormalization group calculations on
  relative energies of transition metal complexes and clusters}.
\newblock \emph{\bibinfo{journal}{The Journal of Chemical Physics}}
  \textbf{\bibinfo{volume}{128}}, \bibinfo{pages}{014104}
  (\bibinfo{year}{2008}).

\bibitem{Mayerle1975}
\bibinfo{author}{Mayerle, J.~J.}, \bibinfo{author}{Denmark, S.~E.},
  \bibinfo{author}{DePamphilis, B.~V.}, \bibinfo{author}{Ibers, J.~A.} \&
  \bibinfo{author}{Holm, R.~H.}
\newblock \bibinfo{title}{{Synthetic analogs of the active sites of iron-sulfur
  proteins. XI. Synthesis and properties of complexes containing the iron
  sulfide Fe$_2$S$_2$ core and the structures of
  bis[o-xylyl-$\alpha$,$\alpha$'-dithiolato-$\mu$-sulfido-ferrate(III)] and
  bis[p-tolylthiolato-$\mu$-sulfido-ferrate(III)] dianions}}.
\newblock \emph{\bibinfo{journal}{Journal of the American Chemical Society}}
  \textbf{\bibinfo{volume}{97}}, \bibinfo{pages}{1032--1045}
  (\bibinfo{year}{1975}).

\bibitem{Orme-Johnson1973}
\bibinfo{author}{Orme-Johnson, W.~H.}
\newblock \bibinfo{title}{{Iron-Sulfur Proteins: Structure and Function}}.
\newblock \emph{\bibinfo{journal}{Annual Review of Biochemistry}}
  \textbf{\bibinfo{volume}{42}}, \bibinfo{pages}{159--204}
  (\bibinfo{year}{1973}).

\bibitem{holmsynthetic}
\bibinfo{author}{{Venkateswara Rao}, P.} \& \bibinfo{author}{Holm, R.~H.}
\newblock \bibinfo{title}{{Synthetic analogues of the active sites of iron
  sulfur proteins}}.
\newblock \emph{\bibinfo{journal}{Chemical Reviews}}
  \textbf{\bibinfo{volume}{104}}, \bibinfo{pages}{527--560}
  (\bibinfo{year}{2003}).

\bibitem{noodleman1996valence}
\bibinfo{author}{Noodleman, L.}, \bibinfo{author}{Case, D.},
  \bibinfo{author}{Mouesca, J.-M.} \& \bibinfo{author}{Lamotte, B.}
\newblock \bibinfo{title}{Valence electron delocalization in polynuclear
  iron-sulfur clusters}.
\newblock \emph{\bibinfo{journal}{JBIC Journal of Biological Inorganic
  Chemistry}} \textbf{\bibinfo{volume}{1}}, \bibinfo{pages}{177--182}
  (\bibinfo{year}{1996}).

\bibitem{Gillum1976}
\bibinfo{author}{Gillum, W.~O.}, \bibinfo{author}{Frankel, R.~B.},
  \bibinfo{author}{Foner, S.} \& \bibinfo{author}{Holm, R.~H.}
\newblock \bibinfo{title}{{Synthetic analogues of the active sites of
  iron-sulfur proteins. XIII. Further electronic structural relationships
  between the analogues [Fe$_2$S$_2$(SR)$_4$]$^{2-}$ and the active sites of
  oxidized 2Fe-2S* proteins}}.
\newblock \emph{\bibinfo{journal}{Inorganic Chemistry}}
  \textbf{\bibinfo{volume}{15}}, \bibinfo{pages}{1095--1100}
  (\bibinfo{year}{1976}).

\bibitem{noodleman1992}
\bibinfo{author}{Noodleman, L.} \& \bibinfo{author}{Case, D.~A.}
\newblock \bibinfo{title}{{Density-functional theory of spin polarization and
  spin coupling in iron-sulfur clusters}}.
\newblock \emph{\bibinfo{journal}{Adv. Inorg. Chem}}
  \textbf{\bibinfo{volume}{38}}, \bibinfo{pages}{423--470}
  (\bibinfo{year}{1992}).

\bibitem{Averill1973}
\bibinfo{author}{Averill, B.~A.}, \bibinfo{author}{Herskovitz, T.},
  \bibinfo{author}{Holm, R.~H.} \& \bibinfo{author}{Ibers, J.~A.}
\newblock \bibinfo{title}{{Synthetic analogs of the active sites of iron-sulfur
  proteins. II. Synthesis and structure of the
  tetra[mercapto-$\mu$3-sulfido-iron] clusters, [Fe$_4$S$_4$(SR)$_4$]$^{2-}$}}.
\newblock \emph{\bibinfo{journal}{Journal of the American Chemical Society}}
  \textbf{\bibinfo{volume}{95}}, \bibinfo{pages}{3523--3534}
  (\bibinfo{year}{1973}).

\bibitem{Noodleman:2002:1367-5931:259}
\bibinfo{author}{Noodleman, L.}, \bibinfo{author}{Lovell, T.},
  \bibinfo{author}{Liu, T.}, \bibinfo{author}{Himo, F.} \&
  \bibinfo{author}{Torres, R.~A.}
\newblock \bibinfo{title}{{Insights into properties and energetics of
  iron-sulfur proteins from simple clusters to nitrogenase}}.
\newblock \emph{\bibinfo{journal}{Current Opinion in Chemical Biology}}
  \textbf{\bibinfo{volume}{6}}, \bibinfo{pages}{259--273}
  (\bibinfo{year}{2002}).

\bibitem{Shaik2005}
\bibinfo{author}{Shaik, S.}, \bibinfo{author}{Kumar, D.},
  \bibinfo{author}{de~Visser, S.~P.}, \bibinfo{author}{Altun, A.} \&
  \bibinfo{author}{Thiel, W.}
\newblock \bibinfo{title}{{Theoretical Perspective on the Structure and
  Mechanism of Cytochrome P450 Enzymes}}.
\newblock \emph{\bibinfo{journal}{Chemical Reviews}}
  \textbf{\bibinfo{volume}{105}}, \bibinfo{pages}{2279--2328}
  (\bibinfo{year}{2005}).

\bibitem{Wodtke2004}
\bibinfo{author}{{Wodtke}, A.~M.}, \bibinfo{author}{Tully, J.~C.} \&
  \bibinfo{author}{Auerbach, D.~J.}
\newblock \bibinfo{title}{{Electronically non-adiabatic interactions of
  molecules at metal surfaces: Can we trust the Born‚ÄìOppenheimer
  approximation for surface chemistry?}}
\newblock \emph{\bibinfo{journal}{International Reviews in Physical Chemistry}}
  \textbf{\bibinfo{volume}{23}}, \bibinfo{pages}{513--539}
  (\bibinfo{year}{2004}).

\bibitem{Frank-Science}
\bibinfo{author}{Lancaster, K.~M.} \emph{et~al.}
\newblock \bibinfo{title}{{X-ray emission spectroscopy evidences a central
  carbon in the nitrogenase iron-molybdenum cofactor}}.
\newblock \emph{\bibinfo{journal}{Science}} \textbf{\bibinfo{volume}{334}},
  \bibinfo{pages}{974--977} (\bibinfo{year}{2011}).

\bibitem{Spatzal18112011}
\bibinfo{author}{Spatzal, T.} \emph{et~al.}
\newblock \bibinfo{title}{{Evidence for Interstitial Carbon in Nitrogenase FeMo
  Cofactor}}.
\newblock \emph{\bibinfo{journal}{Science}} \textbf{\bibinfo{volume}{334}},
  \bibinfo{pages}{940} (\bibinfo{year}{2011}).

\end{thebibliography}
%\bibliographystyle{naturemag}

\newpage
\noindent {\bf Supplementary Information:} 

[2Fe-2S]: geometry and orbitals, DMRG calculations, local charge
and spin, model Hamiltonians

[4Fe-4S]: geometry and orbitals, DMRG calculations, local charge and spin, 
model Hamiltonians

Tables S1-S12

Figures S1-S10

{ Correspondence and requests for materials should be addressed to G.K.C.}

\end{document}

% Double-space the manuscript.

\baselineskip24pt

% Make the title.
\title{Supplementary Information}

\author
{Sandeep Sharma, Kantharuban Sivalingam, Frank Neese, Garnet Kin-Lic Chan}

\maketitle 

\newcommand{\fereduced}{{[Fe$_2$S$_2$(SCH$_3$)$_4$]$^{3-}$} }
\newcommand{\fedimer}{{[Fe$_2$S$_2$(SCH$_3$)$_4$]}}
\newcommand{\fecubane}{{[Fe$_4$S$_4$(SCH$_3$)$_4$]$^{2-}$}}

\tableofcontents
%\listoffigures
%\listoftables

\section{[2Fe-2S] complexes}

\subsection{Geometry and orbitals}

We used a [2Fe-2S] dimer obtained from the
complex of Mayerle {\it et al} \cite{Mayerle1975} by substituting the four terminal toluene groups with methyl groups.
For the \fedimer$^{2-}$ complex, the geometry was derived from the experimental structure reported in \cite{Mayerle1975}, as shown in Supplementary Table \ref{tab:[fes2-geom]}.
For the \fedimer$^{3-}$ complex, we considered two geometries: (i) the geometry in Supplementary Table \ref{tab:[fes2-geom]} (the same geometry as the \fedimer$^{2-}$ complex), 
and (ii) a relaxed geometry, shown in Supplementary Table \ref{tab:[fes3-geom]}. The relaxed geometry was obtained from a broken-symmetry DFT calculation on the $S_z$=1/2 state,
using the BP86 functional and a split-valence with polarization (SVP)  basis set \cite{svbasis} (denoted BP86/SVP) as implemented in Orca\cite{Neese2012}. As seen from the table, in the relaxed geometry
the dimer becomes slightly asymmetric, with the bridging S atoms attracted towards
one of the Fe atoms.

\begin{supptable} 
\caption{Coordinates (in \AA) of the \fedimer$^{2-}$ and unrelaxed \fedimer$^{3-}$ model complexes. \label{tab:[fes2-geom]}}
\begin{center}
\begin{tabular}{lcccc}
\hline
\hline
&& x & y & z\\
\hline
1 &Fe& 5.22& 1.05& -7.95\\
2 &S& 3.86& -0.28& -9.06\\
3 &S& 5.00& 0.95& -5.66\\
4 &S& 4.77& 3.18& -8.74\\
5 &S& 7.23& 0.28& -8.38\\
6 &Fe& 5.88& -1.05& -9.49\\
7 &S& 6.10& -0.95& -11.79\\
8 &S& 6.33& -3.18& -8.71\\
9 &C& 6.00& 4.34& -8.17\\
10 &H& 6.46& 4.81& -9.01\\
11 &H& 5.53& 5.08& -7.55\\
12 &H& 6.74& 3.82& -7.60\\
13 &C& 3.33& 1.31& -5.18\\
14 &H& 2.71&  0.46& -5.37 \\
15 &H& 3.30&  1.54& -4.13 \\
16 &H& 2.97&  2.15& -5.73 \\
17 &C& 5.10& -4.34& -9.28 \\
18 &H& 5.56& -5.05& -9.93 \\
19 &H& 4.67& -4.84& -8.44 \\
20 &H& 4.34& -3.81& -9.81 \\
21 &C& 7.77& -1.31& -12.27\\
22 &H& 7.84& -1.35& -13.34\\
23 &H& 8.42& -0.54& -11.90\\
24 &H& 8.06& -2.25& -11.86\\
\hline
\hline
\end{tabular}
\end{center}
\end{supptable}

\begin{supptable}
\caption{Coordinates (in \AA) of the relaxed \fedimer$^{3-}$ model complex. \label{tab:[fes3-geom]}}
\begin{center}
\begin{tabular}{lcccc}
\hline
\hline
&& x & y & z\\
\hline
1& Fe& 5.48& 1.15& -8.03\\
2 &S& 4.05& -0.61& -8.75\\
3 &S& 5.47& 1.25& -5.58\\
4 &S& 4.63& 3.28& -8.77\\
5 &S& 7.49& 0.42& -9.04\\
6 &Fe& 6.04& -1.22& -9.63\\
7 &S& 5.75& -1.50& -12.05\\
8 &S& 6.86& -3.41& -8.86\\
9 &C& 5.51& 4.45& -7.51\\
10 &H& 6.49& 4.83& -7.92\\
11 &H& 4.87& 5.33& -7.25\\
12 &H& 5.72& 3.84& -6.59\\
13 &C& 3.60& 1.70& -5.54\\
14 &H& 3.01& 0.80& -5.82\\
15 &H& 3.28& 2.06& -4.52\\
16 &H& 3.42& 2.48& -6.31\\
17 &C& 5.21& -4.22& -9.46\\
18 &H& 5.10& -4.01& -10.55\\
19 &H& 5.21& -5.32& -9.26\\
20 &H& 4.37& -3.72& -8.93\\
21 &C& 7.63& -1.85& -12.24\\
22 &H& 7.90& -2.06& -13.31\\
23 &H& 8.20& -0.96& -11.86\\
24 &H& 7.89& -2.72& -11.59\\
\hline
\hline
\end{tabular}
\end{center}
\end{supptable}

To generate the active space for the DMRG calculations, we 
performed an unrestricted DFT BP86/SVP calculation for the high spin ($S_z$=5) state.
The alpha occupied and unoccupied orbitals were then separately localized (``split-localized'') \cite{bytautas} using the Pipek-Mezey algorithm \cite{pipek}.  
From the localized orbitals, iron 3d, 4s, 4d and sulfur 3p orbitals  were identified by 
visual inspection. Some of these orbitals are shown in Supplementary Figure \ref{fig:fedimerorbs}. 

\begin{suppfigure}
\begin{center}
\resizebox{100mm}{!}{\includegraphics{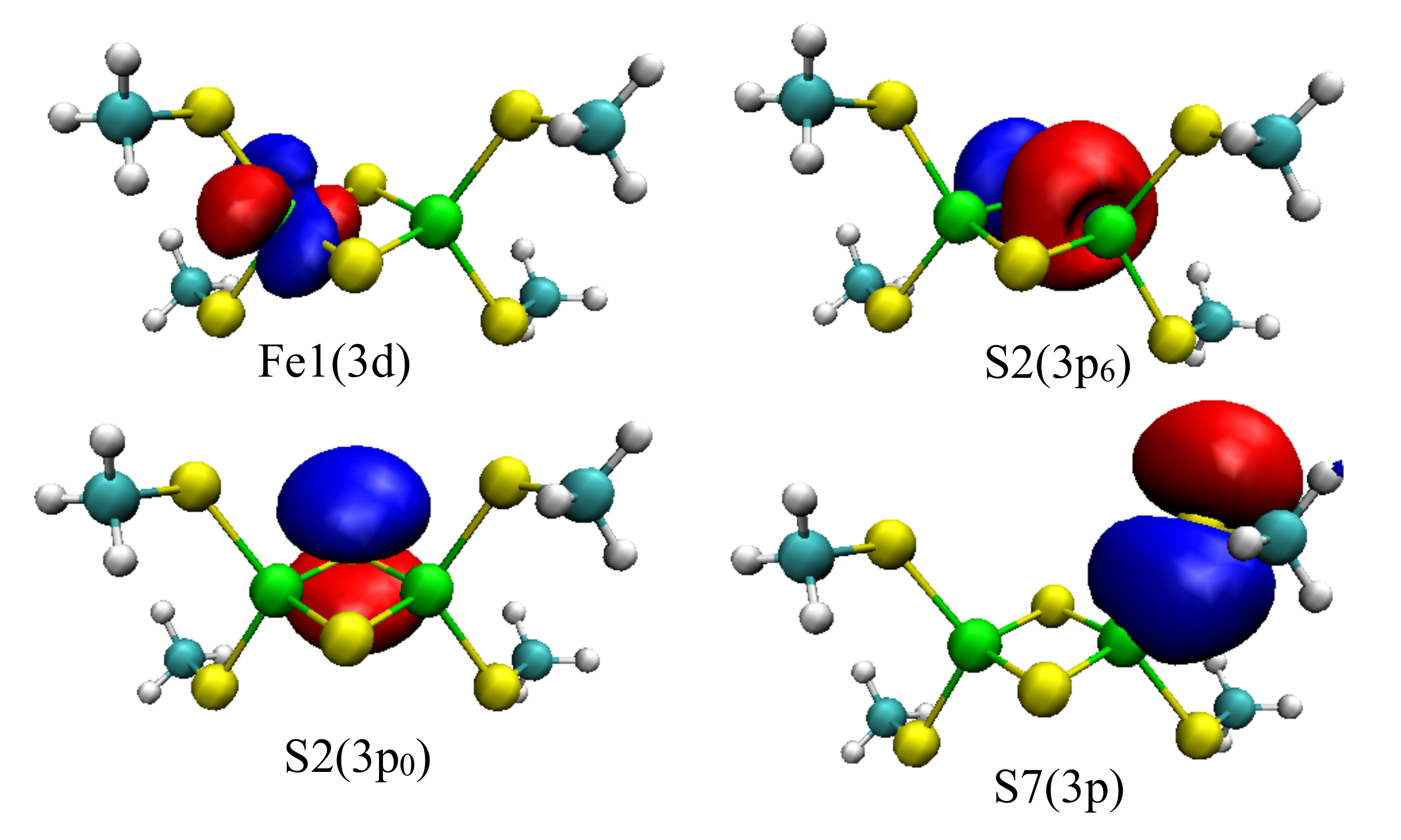}}
\end{center}
\caption{Orbitals in the active space of the [2Fe-2S] dimers.}
\label{fig:fedimerorbs}
\end{suppfigure}

\subsection{DMRG calculations}

\subsubsection{Active spaces}

Four types of active space DMRG calculations (labelled (1)-(4))  were performed on the {\fedimer$^{2-}$} complex
to assess the effect of active space choice. For the
{\fedimer$^{3-}$} complex, we used only active spaces (1) and (2), following the analysis in section \ref{sec:activechoice}.

All DMRG calculations were spin-adapted, using the \textsc{Block} code as described in Ref. \cite{sharma2012}. Thus all
states obtained are eigenfunctions of $S_z$ and $S^2$, and $M$ refers to the
number of spin-adapted renormalized states (the tensor link dimension
in the one-dimensional tensor network underlying the DMRG) which corresponds to effectively twice the
number of non-spin-adapted renormalized states in a standard DMRG calculation \cite{sharma2012}. The 4 types of DMRG calculations were:
\begin{enumerate}
\item
DMRG-CI on a (30e, 20o) active space, with a maximum
of $M$=3500 renormalized states. The 20 orbitals 
included Fe 3d, bridging S 3p, and one 3p orbital per terminal ligand S atom. 
This corresponds to a minimal full valence
active space. For rapid convergence of the DMRG energy, the
orbitals were ordered as follows: S4(3p), S3(3p), Fe1(3d), Fe1(3d), Fe1(3d), Fe1(3d), Fe1(3d),
S2(3p$_1$), S5(3p$_1$), S2(3p$_0$), S5(3p$_0$), S2(3p$_6$), S5(3p$_6$), 
Fe6(3d), Fe6(3d), Fe6(3d), Fe6(3d), Fe6(3d),
S7(3p), S8(3p), where the atom labels 
correspond to the labels in Supplementary Tables \ref{tab:[fes2-geom]} and \ref{tab:[fes3-geom]},
and Figure \ref{fig:geom}, and the subscript on S 3p orbitals is the index of the atom they are pointing towards, with the exception that the 
subscript 0 means that it is pointing in the up-down direction as shown in Figure~\ref{fig:fedimerorbs}.
 \item 
DMRG-CI on a (30e, 32o) active space, with a maximum of $M$=4500 renormalized states.
The 30 orbitals include Fe 4d and Fe 4s orbitals in addition to the 20 described in active space (1). The Fe 4d and 4s orbitals are expected
to account for the principal dynamic and orbital relaxation contributions to the 
energy (i.e. double-shell correlation \cite{Andersson1992}).
The orbitals were ordered as for the 20 orbital active space, with additional Fe 4s and Fe 4d orbitals placed in that order immediately following the 3d orbitals of the same Fe atom.
\item DMRG-SCF on the (30e, 20o) active space (cf. active space(1)). The active space orbitals were optimized using a self-consistent cycle. The DMRG calculations in the SCF optimization used $M$=2500 states. Subsequently a final calculation with $M$=3500 states,
using the fixed  optimized orbitals, was performed. The same
ordering as in active space (1) was used.
\item DMRG-SCF on the (30e, 32o) active space (cf. active space (2)). Again, the DMRG calculations in the SCF optimization were performed with $M$=2500. Subsequently a final calculation with $M$=4500 states, using the fixed optimized orbitals, was performed.
The same ordering as in active space (2) was used.
\end{enumerate}

\subsubsection{Energy convergence}

The DMRG energies and discarded weights at different 
values of $M$ can be used to extrapolate the energy to the $M=\infty$ (FCI) 
result, which corresponds to zero discarded weight. This also
 provides error estimates for 
the DMRG energy \cite{chan2003}. Extrapolations for state-specific DMRG-CI (active space 2) calculations 
are shown in Supplementary Table \ref{tab:dimerextrap} and Supplementary Figure \ref{fig:dimerextrap}. We find that the extrapolated (30e, 32o) 
 singlet and triplet relative energies are converged to within 
0.1 m$E_h$ of the FCI energy.
\begin{supptable}
\caption{DMRG energy in $E_h$ versus the discarded weight of the singlet and triplet states of the \fedimer$^{-2}$ cluster (active space (2), (30e, 32o)). \label{tab:dimerextrap}}
\begin{center}
\begin{tabular} {cccccc}
\hline
\hline
&\multicolumn{2}{c}{Singlet}&&\multicolumn{2}{c}{Triplet}\\
\cline{2-3}\cline{5-6}
M&Discarded weight& Energy&&Discarded weight& Energy\\
\hline
1500	&2.45$\times 10^{-5}$&	-5,104.138933&&2.54$\times 10^{-5}$&	-5,104.135801\\
2500	&1.14$\times 10^{-5}$&	-5,104.139978&&1.23$\times 10^{-5}$&	-5,104.137651\\
3500	&5.63$\times 10^{-6}$&	-5,104.140297&&8.54$\times 10^{-6}$&	-5,104.138315\\
4500	&3.60$\times 10^{-6}$&	-5,104.140426&&6.03$\times 10^{-6}$&	-5,104.138616\\
$\infty$&&-5,104.140718&&&-5,104.139510\\
\hline
\end{tabular}
\end{center}
\end{supptable}

\begin{suppfigure}
\begin{center}
\resizebox{100mm}{!}{\includegraphics{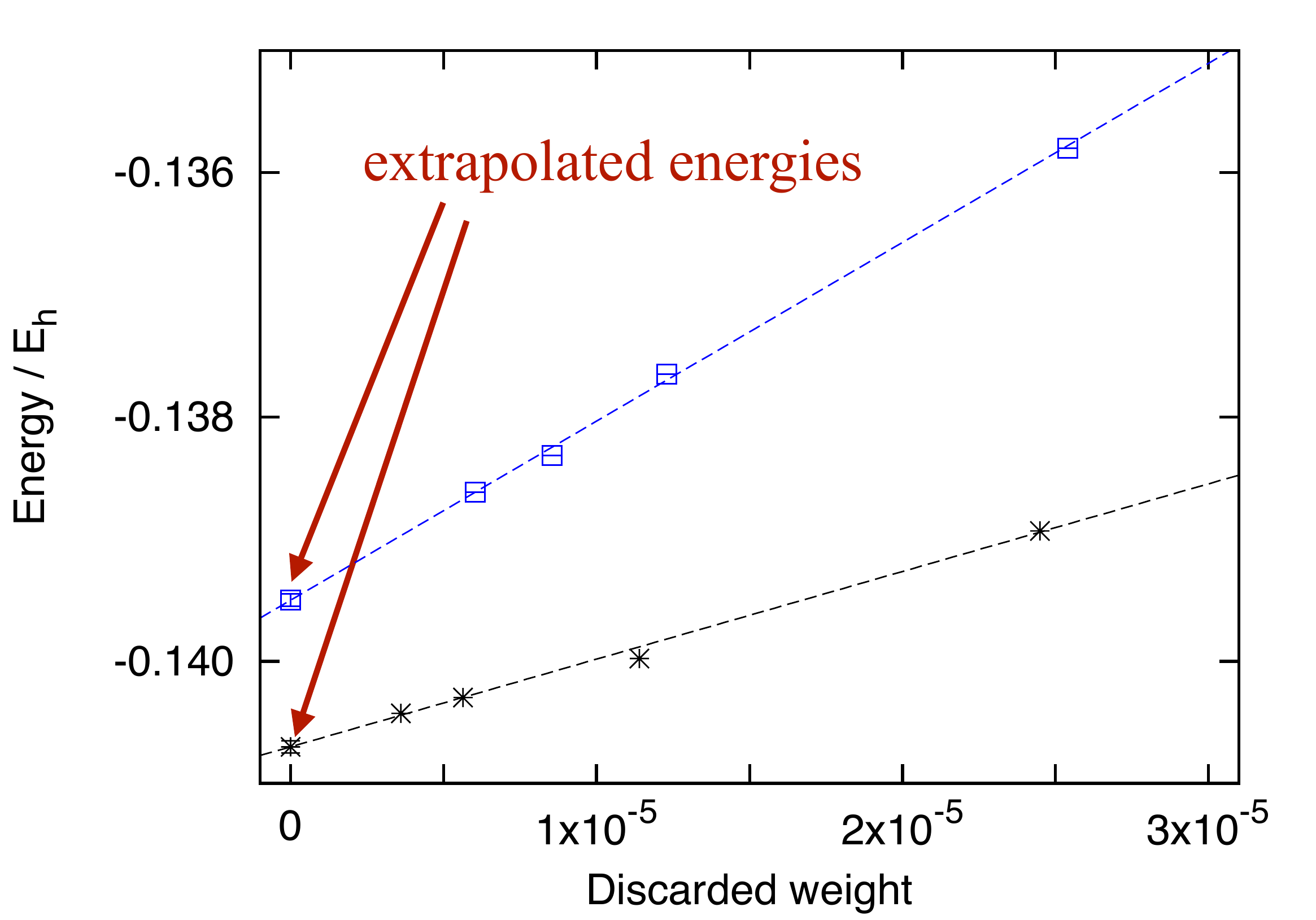}}
\end{center}
\caption{DMRG energy (E+5104.0) in $E_h$ of the singlet and triplet states versus the discarded weight of the \fedimer$^{-2}$ cluster (active space (2), (30e, 32o). The black crosses and the blue dots are respectively the DMRG singlet and triplet state energies and the corresponding lines are the best fit straight lines which are extrapolated to zero discarded weight to obtain an estimated FCI energy.}
\label{fig:dimerextrap}
\end{suppfigure}

For the [2Fe-2S] spectrum calculations, we computed the lowest 10 states 
in each spin-sector
using a state-averaged DMRG calculation. 
%% These calculations are less accurate than the state-specific calculations discussed above, and the
%%  DMRG-CI (30e, 32o) state-averaged relative energies are converged to within [X] m$E_h$ of the exact result. 
Although the energies are not as well converged as for the state-specific calculations, the residual errors
do not qualitatively affect the spectrum or the conclusions of our analysis.

\subsubsection{Assessment of active space}

\label{sec:activechoice}

From the singlet-triplet gap in the active spaces (1)-(4), shown in Supplementary Table \ref{tab:dimerstgap}, we can assess
the effect of the active space choice on the computed energy levels.
We first observe that all 4 active spaces agree closely; even
the minimal valence active space yields a reasonable gap. This
is because the principle exchange pathway leading to the singlet-triplet gap
is via the bridging S 3p ligand orbitals, which are included in the minimal 
active space.
The effect of double-shell correlation in the larger active spaces
is to increase the gap by less than 1.0m$E_h$, while the effect of orbital optimization is very small (0.1m$E_h$ in the larger active space). We take active space (2) (double shell correlation but no orbital optimization) as a practical compromise between accuracy and efficiency. 

\begin{supptable}
\caption{The singlet-triplet gap of the \fedimer$^{2-}$ complex
obtained in active spaces (1)-(4). \label{tab:dimerstgap}}
\begin{center}
\begin{tabular}{ccc}
\hline
\hline
Method & Active Space & Gap/m$E_h$ \\
\hline
DMRG-CI &(30e, 20o) & 1.5 \\
DMRG-CI &(30e, 32o) & 2.1 \\
DMRG-SCF &(30e, 20o)& 1.2 \\ 
DMRG-SCF &(30e, 32o)& 2.0\\
\hline
\hline
\end{tabular}
\end{center}
\end{supptable}

\subsection{Local charge and spin}

%% We define the total spin on an atom $A$ as $\langle S_A^2\rangle$, where the operator
%% is defined as
%% % is this correct, is there the sum over j?
%% \begin{equation}
%% S_A^2 = \sum_{i\in A} [s^2_{zi} + \frac{1}{2}\sum_j (s^+_{i}s^-_j+s^-_{i}s^+_j)]
%% \end{equation}
%% where $i$ labels an orbital on  atom $A$ and $j$ ranges over
%% all orbitals. Note that $\sum_A S_A^2 = S^2$.
To identify the distribution of electrons, we have computed local populations on the atoms.
For atom $A$, the local population $N_A$ is
\begin{equation}
N_A = \sum_{i\in A} n_i \label{eq:localpop}
\end{equation}
where $n_i$ is the number operator of localized orbital $i$ on the $A$.

Further, since our states are eigenstates of $S^2$, there is no spin-density
in the singlet states.
Thus, we have computed local spins and
spin-correlation functions to characterize the electronic structure.
The spin-correlation function between atoms $A$ and $B$, $\langle S_A \cdot S_B\rangle$, is defined as~\cite{Ramos2012,Yamaguchi197735,Yamaguchi1978117}
\begin{eqnarray}
S_A \cdot S_B &=& \sum_\alpha S^\alpha_{A} S^\alpha_B \nonumber\\
S^\alpha_{A} &=& \sum_{i \in A} s^\alpha_{i} \label{eq:spincorr}
\end{eqnarray}
where $\alpha\in \{x,y,z\}$. The local total spin on atom $A$ is defined as
$\langle S_A \cdot S_A \rangle$.

Supplementary Tables \ref{tab:3-s2-2}-\ref{tab:3-s2-relaxed-b} show the relative energies, local populations, spins, and spin-correlation functions for the [2Fe-2S] complexes. All the relative energies reported in the tables are calculated using active space (2). For both geometries state averaged DMRG calculations are performed for the first 10 states with a largest M of 4500. In the case of unrelaxed geometries three sweeps with M=4500 were performed and then its value was reduced in steps of 1000 down to M=1500 to generate reliable extrapolated energies. We find that even though the extrapolation process improves the absolute energies, the energy differences reported in the tables are relatively unchanged. We expect to see the same trend for the relaxed geometries and thus forego the expensive extrapolation step to report the relative DMRG energies calculated with M=4500. 

In Supplementary Table \ref{tab:3-s2-2} and Supplementary Table \ref{tab:3-s2-relaxed} only the 3d, 4s and 4d orbitals of a Fe atom are included in the summations in Equations~\ref{eq:localpop} and \ref{eq:spincorr} to calculate the local electron density and electron spin; whereas in Supplementary Table \ref{tab:3-s2-2-b} and Supplementary Table \ref{tab:3-s2-relaxed-b} the first 16 orbitals and second 16 orbitals (see orbital ordering in previous section for the orbitals) are included in the summations for first and the second Fe atoms respectively.   
The ideal Fe$^{\mathrm{II}}$ and Fe$^{\mathrm{III}}$ populations are 6 and 5 respectively, while the ideal $S$=2 and $S$=5/2 total spins are 6 and 8.75 respectively.
We see that the observed local populations and total spins are increased and reduced respectively in the complexes
due to the effect of quantum fluctuations, such as delocalization onto adjacent sulfur orbitals.
We also see from the spin-correlation functions that the spins progressively move from being anti-aligned to aligned as the
the total dimer spin is increased. 
\begin{supptable}
\caption{The local population, total spin, and spin-correlation functions in the lowest spin states of the {\fedimer$^{2-}$} complex,
using active space (2). Note $\langle N_2\rangle=\langle N_1\rangle$ and $\langle S_2^2\rangle =\langle S_1^2\rangle$. One orbitals of the Fe
\label{tab:2-s2-2}}
\begin{center}
\begin{tabular}{ccccc}
\hline
\hline
Dimer $S$ & $\langle N_1\rangle$ & $\langle S_1^2\rangle$ & $\langle S_1 \cdot S_2\rangle$ \\
\hline
0 & 6.18&5.47 &-4.92  \\
1 & 6.17 &5.47&-4.22  \\
2 & 6.17&5.49 &-2.84  \\
3 & 6.17&5.51 & -0.79 \\
4 & 6.16&5.54 &1.84  \\
5 & 6.13&5.74 & 3.74 \\
\hline
\hline
\end{tabular}
\end{center}
\end{supptable}

\subsubsection{Asymmetry in the \fedimer$^{3-}$ complex}

The relaxed geometry of the \fedimer$^{3-}$ complex is slightly asymmetric. The local populations and spins in Supplementary Table
\ref{tab:3-s2-relaxed} and \ref{tab:3-s2-relaxed-b} show the effect of this asymmetry on the electronic structure.
As observed in Supplementary Table~\ref{tab:3-s2-relaxed} the asymmetry in the Fe atoms appears rather small, amounting to
up to 0.05 electron units in the population, and 0.2 spin units in the 
local spin. But when we compare the local spins shown in Supplementary Table~\ref{tab:3-s2-relaxed-b}, the asymmetry is much larger with differences between the local spins as large as 2.4 in some cases. The difference in asymmetry between the two tables points to the fact that the asymmetry is largely due to the difference in the S 3p orbitals.

\begin{supptable}
\caption{Unrelaxed geometry {\fedimer$^{3-}$}  energy (cm$^{-1}$), local population, total spin and spin-correlation functions for the lowest ten levels in each  dimer total spin state,
using active space (2).\label{tab:3-s2-2}}
\begin{center}
\begin{tabular}{ccccccccccc}
\hline
\hline
State   & 1 & 2 & 3 & 4 & 5 & 6 & 7 & 8 & 9 & 10\\
\hline
$S$=1/2 &   &   &   &   &   &   &   &   &   & \\
\hline
$E$&0	&325	&1132&	2642&	4264&	4989&	4905&5313&	6448&7049\\
$\langle N_1 \rangle$ &   6.23 &   6.25 &   6.24 &   6.23 &   6.22 &   6.21 &   6.22 &   6.21 &   6.21 &   6.21 \\
$\langle N_2 \rangle$ &    6.24 &    6.23 &    6.24 &    6.23 &    6.22 &    6.22 &    6.21 &    6.21 &    6.21 &    6.21 \\
$\langle S_1\rangle$ &    5.39 &    5.31 &    5.34 &    5.35 &    5.39 &    5.40 &    5.36 &    5.43 &    5.40 &    5.42 \\
$\langle S_2\rangle$ &    5.32 &    5.36 &    5.34 &    5.35 &    5.36 &    5.37 &    5.43 &    5.42 &    5.41 &    5.42 \\
$\langle S_1\cdot S_2\rangle$ &   -4.74 &   -4.77 &   -4.81 &   -4.62 &   -4.84 &   -4.72 &   -4.83 &   -4.89 &   -4.78 &   -4.77 \\
\hline
$S$=3/2 &   &   &   &   &   &   &   &   &   & \\
\hline
$E$&136&	527	&1710	&4264	&4451	&5030	&5131	&6073	&7870	&8581\\
$\langle N_1 \rangle$ &   6.22 &   6.24 &   6.24 &   6.22 &   6.21 &   6.21 &   6.21 &   6.22 &   6.21 &   6.20 \\
$\langle N_2 \rangle$ &    6.23 &    6.24 &    6.23 &    6.22 &    6.22 &    6.21 &    6.21 &    6.21 &    6.21 &    6.20 \\
$\langle S_1\rangle$ &    5.42 &    5.35 &    5.34 &    5.37 &    5.40 &    5.42 &    5.43 &    5.38 &    5.42 &    5.43 \\
$\langle S_2\rangle$ &    5.39 &    5.36 &    5.36 &    5.37 &    5.40 &    5.42 &    5.43 &    5.41 &    5.42 &    5.43 \\
$\langle S_1\cdot S_2\rangle$ &   -3.66 &   -3.77 &   -3.64 &   -3.38 &   -3.68 &   -3.79 &   -3.78 &   -3.47 &   -3.61 &   -3.59 \\
\hline
$S$=5/2 &   &   &   &   &   &   &   &   &   & \\
\hline
$E$&336	&643	&2871	&4668	&5300	&5678	&6248	&7580	&9459	&10260\\
$\langle N_1 \rangle$ &   6.21 &   6.23 &   6.23 &   6.20 &   6.21 &   6.20 &   6.22 &   6.21 &   6.20 &   6.20 \\
$\langle N_2 \rangle$ &    6.22 &    6.24 &    6.23 &    6.21 &    6.21 &    6.21 &    6.22 &    6.20 &    6.20 &    6.20 \\
$\langle S_1\rangle$ &    5.50 &    5.38 &    5.37 &    5.45 &    5.44 &    5.46 &    5.38 &    5.41 &    5.44 &    5.43 \\
$\langle S_2\rangle$ &    5.45 &    5.37 &    5.38 &    5.43 &    5.44 &    5.44 &    5.39 &    5.45 &    5.45 &    5.44 \\
$\langle S_1\cdot S_2\rangle$ &   -1.89 &   -1.96 &   -1.77 &   -1.81 &   -1.91 &   -1.95 &   -1.33 &   -1.51 &   -1.68 &   -1.67 \\
\hline
$S$=7/2 &   &   &   &   &   &   &   &   &   & \\
\hline
$E$&669	&1330	&4675	&5441	&6358	&7049	&8989	&9589	&9913	&10115\\
$\langle N_1 \rangle$ &   6.19 &   6.22 &   6.22 &   6.19 &   6.19 &   6.20 &   6.20 &   6.25 &   6.22 &   6.28 \\
$\langle N_2 \rangle$ &    6.20 &    6.23 &    6.22 &    6.19 &    6.21 &    6.19 &    6.21 &    6.27 &    6.24 &    6.30 \\
$\langle S_1\rangle$ &    5.56 &    5.43 &    5.43 &    5.50 &    5.51 &    5.50 &    5.45 &    5.02 &    5.29 &    4.76 \\
$\langle S_2\rangle$ &    5.55 &    5.42 &    5.44 &    5.51 &    5.46 &    5.54 &    5.40 &    5.00 &    5.25 &    4.76 \\
$\langle S_1\cdot S_2\rangle$ &    0.55 &    0.60 &    0.78 &    0.78 &    0.67 &    0.61 &    1.38 &    2.16 &    1.72 &    2.49 \\
\hline
$S$=9/2 &   &   &   &   &   &   &   &   &   & \\
\hline
$E$&1071	&2943	&7357	&7403	&8496	&9209	&13263	&13699	&14803	&15073\\
$\langle N_1 \rangle$ &   6.18 &   6.21 &   6.16 &   6.20 &   6.19 &   6.17 &   6.17 &   6.17 &   6.17 &   6.22 \\
$\langle N_2 \rangle$ &    6.19 &    6.21 &    6.16 &    6.20 &    6.18 &    6.17 &    6.17 &    6.17 &    6.17 &    6.22 \\
$\langle S_1\rangle$ &    5.63 &    5.57 &    5.68 &    5.59 &    5.60 &    5.64 &    5.65 &    5.65 &    5.64 &    5.53 \\
$\langle S_2\rangle$ &    5.62 &    5.56 &    5.67 &    5.58 &    5.62 &    5.65 &    5.66 &    5.65 &    5.64 &    5.53 \\
$\langle S_1\cdot S_2\rangle$ &    3.62 &    3.58 &    3.67 &    3.60 &    3.62 &    3.65 &    3.66 &    3.65 &    3.65 &    3.56 \\
\hline
\hline
\end{tabular}
\end{center}
\end{supptable}

\begin{supptable}
\caption{Unrelaxed geometry {\fedimer$^{3-}$}  energy (cm$^{-1}$), local population, total spin and spin-correlation functions for the lowest ten levels in each  dimer total spin state,
using active space (2). The orbitals taken to \label{tab:3-s2-2-b}}
\begin{center}
\begin{tabular}{ccccccccccc}
\hline
\hline
State   & 1 & 2 & 3 & 4 & 5 & 6 & 7 & 8 & 9 & 10\\
\hline
$S$=1/2 &   &   &   &   &   &   &   &   &   & \\
\hline
$E$&0	&325	&1132&	2642&	4264&	4989&	4905&5313&	6448&7049\\
$\langle N_1 \rangle$ &  15.48 &  15.50 &  15.49 &  15.49 &  15.49 &  15.48 &  15.52 &  15.49 &  15.50 &  15.49 \\
$\langle N_2 \rangle$ &   15.52 &   15.50 &   15.51 &   15.51 &   15.51 &   15.52 &   15.48 &   15.51 &   15.50 &   15.51 \\
$\langle S_1\rangle$ &    5.87 &    5.63 &    5.77 &    5.64 &    5.84 &    5.78 &    5.62 &    5.84 &    5.74 &    5.76 \\
$\langle S_2\rangle$ &    5.61 &    5.84 &    5.76 &    5.68 &    5.63 &    5.59 &    5.95 &    5.83 &    5.79 &    5.78 \\
$\langle S_1\cdot S_2\rangle$ &   -5.37 &   -5.36 &   -5.39 &   -5.28 &   -5.36 &   -5.31 &   -5.41 &   -5.46 &   -5.39 &   -5.40 \\
\hline
$S$=3/2 &   &   &   &   &   &   &   &   &   & \\
\hline
$E$&136&	527	&1710	&4264	&4451	&5030	&5131	&6073	&7870	&8581\\
$\langle N_1 \rangle$ &  15.49 &  15.50 &  15.50 &  15.49 &  15.50 &  15.49 &  15.49 &  15.50 &  15.50 &  15.49 \\
$\langle N_2 \rangle$ &   15.51 &   15.50 &   15.50 &   15.51 &   15.50 &   15.51 &   15.51 &   15.50 &   15.50 &   15.51 \\
$\langle S_1\rangle$ &    6.00 &    5.95 &    5.85 &    5.76 &    5.93 &    6.00 &    5.99 &    5.74 &    5.89 &    5.90 \\
$\langle S_2\rangle$ &    5.89 &    5.98 &    5.92 &    5.78 &    5.86 &    5.95 &    5.98 &    5.84 &    5.91 &    5.88 \\
$\langle S_1\cdot S_2\rangle$ &   -4.07 &   -4.09 &   -4.01 &   -3.90 &   -4.02 &   -4.10 &   -4.11 &   -3.91 &   -4.03 &   -4.01 \\
\hline
$S$=5/2 &   &   &   &   &   &   &   &   &   & \\
\hline
$E$&336	&643	&2871	&4668	&5300	&5678	&6248	&7580	&9459	&10260\\
$\langle N_1 \rangle$ &  15.49 &  15.49 &  15.50 &  15.50 &  15.49 &  15.49 &  15.49 &  15.50 &  15.50 &  15.49 \\
$\langle N_2 \rangle$ &   15.51 &   15.51 &   15.50 &   15.50 &   15.51 &   15.51 &   15.51 &   15.50 &   15.50 &   15.51 \\
$\langle S_1\rangle$ &    6.35 &    6.27 &    6.13 &    6.24 &    6.24 &    6.32 &    5.91 &    5.95 &    6.11 &    6.12 \\
$\langle S_2\rangle$ &    6.22 &    6.25 &    6.17 &    6.18 &    6.26 &    6.16 &    5.94 &    6.10 &    6.16 &    6.13 \\
$\langle S_1\cdot S_2\rangle$ &   -1.91 &   -1.89 &   -1.78 &   -1.84 &   -1.87 &   -1.87 &   -1.55 &   -1.65 &   -1.76 &   -1.75 \\
\hline
$S$=7/2 &   &   &   &   &   &   &   &   &   & \\
\hline
$E$&669	&1330	&4675	&5441	&6358	&7049	&8989	&9589	&9913	&10115\\
$\langle N_1 \rangle$ &  15.49 &  15.49 &  15.50 &  15.50 &  15.48 &  15.50 &  15.48 &  15.46 &  15.46 &  15.54 \\
$\langle N_2 \rangle$ &   15.51 &   15.51 &   15.50 &   15.50 &   15.52 &   15.50 &   15.52 &   15.54 &   15.54 &   15.46 \\
$\langle S_1\rangle$ &    6.76 &    6.67 &    6.55 &    6.56 &    6.70 &    6.62 &    6.31 &    5.59 &    5.98 &    5.46 \\
$\langle S_2\rangle$ &    6.78 &    6.69 &    6.61 &    6.67 &    6.62 &    6.76 &    6.19 &    5.65 &    5.97 &    5.21 \\
$\langle S_1\cdot S_2\rangle$ &    1.10 &    1.19 &    1.30 &    1.26 &    1.22 &    1.19 &    1.63 &    2.26 &    1.90 &    2.54 \\
\hline
$S$=9/2 &   &   &   &   &   &   &   &   &   & \\
\hline
$E$&1071	&2943	&7357	&7403	&8496	&9209	&13263	&13699	&14803	&15073\\
$\langle N_1 \rangle$ &  15.49 &  15.49 &  15.50 &  15.49 &  15.50 &  15.49 &  15.48 &  15.51 &  15.49 &  15.50 \\
$\langle N_2 \rangle$ &   15.51 &   15.51 &   15.50 &   15.51 &   15.50 &   15.51 &   15.52 &   15.49 &   15.51 &   15.50 \\
$\langle S_1\rangle$ &    7.44 &    7.43 &    7.41 &    7.45 &    7.42 &    7.44 &    7.46 &    7.39 &    7.44 &    7.42 \\
$\langle S_2\rangle$ &    7.40 &    7.40 &    7.40 &    7.38 &    7.41 &    7.39 &    7.37 &    7.44 &    7.38 &    7.40 \\
$\langle S_1\cdot S_2\rangle$ &    4.95 &    4.96 &    4.97 &    4.96 &    4.96 &    4.96 &    4.96 &    4.96 &    4.96 &    4.96 \\
\hline
\hline
\end{tabular}
\end{center}
\end{supptable}

\begin{supptable}
\caption{Relaxed geometry {\fedimer$^{3-}$} energy (cm$^{-1}$), local population, total spin and spin-correlation functions for the lowest ten levels in each  dimer total spin state,
using active space (2).\label{tab:3-s2-relaxed}}
\begin{center}
\begin{tabular}{ccccccccccc}
\hline
\hline
State   & 1 & 2 & 3 & 4 & 5 & 6 & 7 & 8 & 9 & 10\\
\hline
$S$=1/2 &   &   &   &   &   &   &   &   &   & \\
\hline
$E$&0	&1218	&2079	&3790	&4070	&4314	&4885	&5731	&6975	&7644\\
$\langle N_1 \rangle$ &   6.26 &   6.24 &   6.16 &   6.18 &   6.21 &   6.21 &   6.20 &   6.17 &   6.17 &   6.17 \\
$\langle N_2 \rangle$ &    6.14 &    6.16 &    6.28 &    6.25 &    6.16 &    6.16 &    6.16 &    6.23 &    6.22 &    6.22 \\
$\langle S_1\rangle$ &    5.32 &    5.37 &    5.67 &    5.63 &    5.45 &    5.42 &    5.48 &    5.62 &    5.65 &    5.64 \\
$\langle S_2\rangle$ &    5.73 &    5.67 &    5.18 &    5.26 &    5.68 &    5.67 &    5.65 &    5.33 &    5.34 &    5.31 \\
$\langle S_1\cdot S_2\rangle$ &   -4.96 &   -5.01 &   -4.85 &   -4.78 &   -5.03 &   -5.04 &   -5.02 &   -4.86 &   -4.90 &   -4.89 \\
\hline
$S$=3/2 &   &   &   &   &   &   &   &   &   & \\
\hline
$E$&266	&1211	&2374	&4197	&4493	&4694	&4925	&6460	&7693	&8342\\
$\langle N_1 \rangle$ &   6.24 &   6.23 &   6.17 &   6.20 &   6.19 &   6.18 &   6.19 &   6.17 &   6.17 &   6.24 \\
$\langle N_2 \rangle$ &    6.15 &    6.17 &    6.26 &    6.18 &    6.19 &    6.21 &    6.16 &    6.22 &    6.21 &    6.28 \\
$\langle S_1\rangle$ &    5.37 &    5.42 &    5.64 &    5.50 &    5.53 &    5.57 &    5.49 &    5.61 &    5.63 &    5.37 \\
$\langle S_2\rangle$ &    5.72 &    5.63 &    5.28 &    5.61 &    5.51 &    5.44 &    5.67 &    5.39 &    5.39 &    4.93 \\
$\langle S_1\cdot S_2\rangle$ &   -3.86 &   -3.97 &   -3.75 &   -3.90 &   -3.74 &   -3.67 &   -3.89 &   -3.61 &   -3.72 &   -2.71 \\
\hline
$S$=5/2 &   &   &   &   &   &   &   &   &   & \\
\hline
$E$&623	&1323	&2848	&4619	&4913	&5200	&5943	&7536	&8374	&8715\\
$\langle N_1 \rangle$ &   6.23 &   6.22 &   6.18 &   6.18 &   6.18 &   6.19 &   6.18 &   6.17 &   6.24 &   6.18 \\
$\langle N_2 \rangle$ &    6.15 &    6.18 &    6.22 &    6.20 &    6.17 &    6.16 &    6.23 &    6.21 &    6.28 &    6.21 \\
$\langle S_1\rangle$ &    5.44 &    5.47 &    5.59 &    5.57 &    5.55 &    5.52 &    5.60 &    5.62 &    5.34 &    5.61 \\
$\langle S_2\rangle$ &    5.70 &    5.62 &    5.44 &    5.52 &    5.61 &    5.67 &    5.38 &    5.43 &    4.98 &    5.41 \\
$\langle S_1\cdot S_2\rangle$ &   -2.03 &   -2.16 &   -1.96 &   -1.98 &   -2.04 &   -2.03 &   -1.43 &   -1.59 &   -0.24 &   -1.76 \\
\hline
$S$=7/2 &   &   &   &   &   &   &   &   &   & \\
\hline
$E$&786	&1405	&3375	&5173	&5368	&5696	&7528	&8254	&8863	&9149\\
$\langle N_1 \rangle$ &   6.21 &   6.21 &   6.20 &   6.17 &   6.17 &   6.18 &   6.17 &   6.24 &   6.17 &   6.24 \\
$\langle N_2 \rangle$ &    6.15 &    6.18 &    6.18 &    6.17 &    6.21 &    6.16 &    6.22 &    6.27 &    6.20 &    6.27 \\
$\langle S_1\rangle$ &    5.52 &    5.51 &    5.55 &    5.60 &    5.63 &    5.57 &    5.61 &    5.33 &    5.62 &    5.33 \\
$\langle S_2\rangle$ &    5.70 &    5.62 &    5.61 &    5.62 &    5.52 &    5.67 &    5.42 &    5.03 &    5.47 &    5.06 \\
$\langle S_1\cdot S_2\rangle$ &    0.50 &    0.39 &    0.53 &    0.60 &    0.60 &    0.52 &    1.33 &    3.07 &    1.18 &    2.98 \\
\hline
$S$=9/2 &   &   &   &   &   &   &   &   &   & \\
\hline
$E$&984	&1839	&4254	&5941	&6458	&6873	&10136	&11196	&11377	&12487\\
$\langle N_1 \rangle$ &   6.19 &   6.20 &   6.20 &   6.16 &   6.17 &   6.15 &   6.15 &   6.15 &   6.18 &   6.16 \\
$\langle N_2 \rangle$ &    6.16 &    6.18 &    6.15 &    6.14 &    6.17 &    6.21 &    6.20 &    6.19 &    6.21 &    6.19 \\
$\langle S_1\rangle$ &    5.60 &    5.57 &    5.56 &    5.68 &    5.66 &    5.71 &    5.70 &    5.70 &    5.64 &    5.69 \\
$\langle S_2\rangle$ &    5.72 &    5.66 &    5.73 &    5.74 &    5.67 &    5.55 &    5.58 &    5.60 &    5.56 &    5.59 \\
$\langle S_1\cdot S_2\rangle$ &    3.66 &    3.63 &    3.66 &    3.71 &    3.68 &    3.65 &    3.66 &    3.67 &    3.63 &    3.66 \\
\hline
\hline
\end{tabular}
\end{center}
\end{supptable}

\begin{supptable}
\caption{Relaxed geometry {\fedimer$^{3-}$} energy (cm$^{-1}$), local population, total spin and spin-correlation functions for the lowest ten levels in each  dimer total spin state,
using active space (2).\label{tab:3-s2-relaxed-b}}
\begin{center}
\begin{tabular}{ccccccccccc}
\hline
\hline
State   & 1 & 2 & 3 & 4 & 5 & 6 & 7 & 8 & 9 & 10\\
\hline
$S$=1/2 &   &   &   &   &   &   &   &   &   & \\
\hline
$E$&0	&1218	&2079	&3790	&4070	&4314	&4885	&5731	&6975	&7644\\
$\langle N_1 \rangle$ &  15.62 &  15.59 &  15.37 &  15.39 &  15.61 &  15.61 &  15.58 &  15.41 &  15.37 &  15.38 \\
$\langle N_2 \rangle$ &   15.38 &   15.41 &   15.63 &   15.61 &   15.39 &   15.39 &   15.42 &   15.59 &   15.63 &   15.62 \\
$\langle S_1\rangle$ &    5.05 &    5.20 &    7.31 &    7.11 &    5.24 &    5.26 &    5.30 &    6.89 &    7.28 &    7.23 \\
$\langle S_2\rangle$ &    7.42 &    7.25 &    5.04 &    5.13 &    7.30 &    7.27 &    7.02 &    5.31 &    5.11 &    5.02 \\
$\langle S_1\cdot S_2\rangle$ &   -5.86 &   -5.85 &   -5.80 &   -5.74 &   -5.89 &   -5.89 &   -5.79 &   -5.72 &   -5.82 &   -5.75 \\
\hline
$S$=3/2 &   &   &   &   &   &   &   &   &   & \\
\hline
$E$&266	&1211	&2374	&4197	&4493	&4694	&4925	&6460	&7693	&8342\\
$\langle N_1 \rangle$ &  15.61 &  15.56 &  15.40 &  15.57 &  15.53 &  15.46 &  15.58 &  15.43 &  15.40 &  15.30 \\
$\langle N_2 \rangle$ &   15.39 &   15.44 &   15.60 &   15.43 &   15.47 &   15.54 &   15.42 &   15.57 &   15.60 &   15.70 \\
$\langle S_1\rangle$ &    5.38 &    5.70 &    7.11 &    5.79 &    5.95 &    6.57 &    5.55 &    6.80 &    7.09 &    6.21 \\
$\langle S_2\rangle$ &    7.38 &    7.10 &    5.51 &    6.96 &    6.43 &    6.05 &    7.07 &    5.53 &    5.47 &    5.11 \\
$\langle S_1\cdot S_2\rangle$ &   -4.51 &   -4.53 &   -4.43 &   -4.50 &   -4.32 &   -4.43 &   -4.44 &   -4.29 &   -4.41 &   -3.78 \\
\hline
$S$=5/2 &   &   &   &   &   &   &   &   &   & \\
\hline
$E$&623	&1323	&2848	&4619	&4913	&5200	&5943	&7536	&8374	&8715\\
$\langle N_1 \rangle$ &  15.60 &  15.55 &  15.45 &  15.51 &  15.54 &  15.59 &  15.42 &  15.43 &  15.29 &  15.41 \\
$\langle N_2 \rangle$ &   15.40 &   15.45 &   15.55 &   15.49 &   15.46 &   15.41 &   15.58 &   15.57 &   15.71 &   15.59 \\
$\langle S_1\rangle$ &    5.90 &    6.19 &    6.83 &    6.45 &    6.15 &    5.85 &    6.72 &    6.88 &    5.55 &    7.01 \\
$\langle S_2\rangle$ &    7.33 &    7.10 &    6.25 &    6.65 &    6.91 &    7.19 &    5.79 &    5.71 &    5.27 &    5.83 \\
$\langle S_1\cdot S_2\rangle$ &   -2.24 &   -2.27 &   -2.16 &   -2.18 &   -2.16 &   -2.15 &   -1.88 &   -1.92 &   -1.03 &   -2.05 \\
\hline
$S$=7/2 &   &   &   &   &   &   &   &   &   & \\
\hline
$E$&786	&1405	&3375	&5173	&5368	&5696	&7528	&8254	&8863	&9149\\
$\langle N_1 \rangle$ &  15.58 &  15.54 &  15.51 &  15.53 &  15.48 &  15.57 &  15.43 &  15.28 &  15.42 &  15.39 \\
$\langle N_2 \rangle$ &   15.42 &   15.46 &   15.49 &   15.47 &   15.52 &   15.43 &   15.57 &   15.72 &   15.58 &   15.61 \\
$\langle S_1\rangle$ &    6.53 &    6.69 &    6.76 &    6.68 &    6.92 &    6.50 &    6.66 &    5.01 &    7.10 &    5.29 \\
$\langle S_2\rangle$ &    7.37 &    7.23 &    7.00 &    7.06 &    6.69 &    7.26 &    6.17 &    5.46 &    5.90 &    5.23 \\
$\langle S_1\cdot S_2\rangle$ &    0.92 &    0.92 &    0.99 &    1.00 &    1.07 &    1.00 &    1.46 &    2.64 &    1.38 &    2.61 \\
\hline
$S$=9/2 &   &   &   &   &   &   &   &   &   & \\
\hline
$E$&984	&1839	&4254	&5941	&6458	&6873	&10136	&11196	&11377	&12487\\
$\langle N_1 \rangle$ &  15.56 &  15.54 &  15.57 &  15.55 &  15.54 &  15.42 &  15.48 &  15.29 &  15.37 &  15.44 \\
$\langle N_2 \rangle$ &   15.44 &   15.46 &   15.43 &   15.45 &   15.46 &   15.58 &   15.52 &   15.71 &   15.63 &   15.56 \\
$\langle S_1\rangle$ &    7.23 &    7.28 &    7.20 &    7.25 &    7.27 &    7.60 &    7.45 &    7.95 &    7.74 &    7.56 \\
$\langle S_2\rangle$ &    7.55 &    7.50 &    7.57 &    7.53 &    7.51 &    7.18 &    7.33 &    6.83 &    7.03 &    7.21 \\
$\langle S_1\cdot S_2\rangle$ &    4.98 &    4.99 &    4.99 &    4.98 &    4.99 &    4.98 &    4.99 &    4.98 &    4.99 &    4.99 \\
\hline
\hline
\end{tabular}
\end{center}
\end{supptable}

\subsection{Model Hamiltonian for the \fereduced dimer}

The energy levels of the HDE model for the \fereduced mixed valence complex, as derived by Noodleman and Baerends,
are given by
\begin{equation}
E(S)=2J S_1 \cdot S_2 \pm B(S+1/2) \label{eq:hdereduced}
\end{equation}
As demonstrated in the main text, the HDE energy levels do not fit the {\it ab-initio} DMRG spectrum well
because of the   assumptions used to derive Eq. (\ref{eq:hdereduced}).
Before deriving a more complete model that {\it is} compatible with the {\it ab-initio} spectrum, we briefly recall how Eq. (\ref{eq:hdereduced})
 is obtained from Anderson's  analysis of double exchange \cite{anderson-DE,Noodleman1984,shoji2007theory2}.

We first consider an oxidized complex with two ferric ions (with spins $S_1=5/2$, $S_2=5/2$)  as a ``base'' system. 
The extra electron in the reduced dimer is added to this base system, where it hops between a pair of local orbitals
on each of the ions. Denoting the creation (annihilation) operators for the local orbitals on the first (second)
ions by $c^{(\dag)}_{1}$, $c^{(\dag)}_{2}$ respectively, and the spin of the electron 
 as $s_1$, $s_2$ respectively, 
Anderson's analysis\cite{anderson-DE} leads to a Hamiltonian of the form 

\begin{equation}
H = J(S_1\cdot S_2+S_1 \cdot s_2 + S_2 \cdot s_1)  + \sum_{\sigma={\uparrow,\downarrow}} \beta(c^\dag_{1\sigma} c_{2\sigma} + 
c^\dag_{2\sigma} c_{1\sigma})
\label{eq:andersondimer}
\end{equation}
where the Hamiltonian is to be solved in the Hilbert space where the hopping electron is always anti-aligned with the spin
of the ferric ion on which it is currently residing. 
%This can formally be achieved by working in the full space
%with a constraint term $I(S_1 \cdot s_1 + S_2 \cdot s_2)$ and taking the limit $I \to \infty$.
The terms in $H$ have the following meaning:
\begin{enumerate}
\item The first corresponds to Heisenberg exchange coupling between  spins on the two ions (the ``base'' ferric spins and
the extra spin of the hopping electron).
%% \item The second is a constraint term to enforce the Pauli principle as the hopping electron must be anti-aligned with the spin of the ferric ion 
%% on which it is currently residing. This can be enforced by taking the limit $I\to\infty$.
\item The second describes the effective hopping of the electron between the two ferric ions (the sum over
$\sigma$ is a summation over electron spin).
\end{enumerate}
The eigenvalues of the Anderson Hamiltonian may be determined analytically to be the HDE energy levels in Eq. (\ref{eq:hdereduced}), where
 $B=\beta/(2S+1)$, $S=S_1=S_2$.
	
As argued in the main text, the most commonly used 
version of the HDE model breaks down in the {\fereduced}
dimer because  it assumes that there is a {\it single} pair of
d orbitals on the ferric ions that participates in the hopping process.
%% , and (ii)
%% it assumes that each ferric ion always remains in the $S=5/2$ state.
%% We discuss   (i) first. 
This assumption is valid if the
double exchange splitting $B(S+1/2)$ is much smaller than the 
ligand-field splitting $\Delta$. However, this is clearly not the case for Fe ions with
tetrahedral coordination which is typically associated with weak ligand fields. 
Instead, all 5 pairs of d orbitals participate in the hopping at low energies.

We can extend Anderson's double exchange Hamiltonian to multi-orbital double exchange. We label each of the 5 local d orbitals by
index $i$. This gives
% Form correct, particularly with Heisenberg exchange.
\begin{eqnarray}
H &=& \sum_{ij} J_{ij} s_{1i} \cdot s_{2j} 
  + \sum_{i\sigma} \left[\beta_i (c_{1i\sigma}^\dag c_{2i\sigma} + c_{2i\sigma}^\dag c_{1i\sigma}) + \Delta_i (c_{1i\sigma}^\dag c_{1i\sigma} + c_{2i\sigma}^\dag c_{2i\sigma})\right]
\label{eq:multibaseh}
\end{eqnarray}
where we once again restrict ourselves to states where the hopping electron is strictly antiferromagnetically aligned to the base spins (the base
spins are all ferromagnetically aligned). 
The additional $\Delta_i$ term gives
the ligand field splitting of the orbitals. The above form has a very large number of parameters from the general exchange couplings $J_{ij}$. 
However, in the limiting case where all spins are aligned on each Fe atom (e.g. as in the oxidized dimer), then it is sufficient to
consider an exchange term of the form $\sum_i J'_i s_{1i} \cdot s_{2i}$ since the interaction of any spin on a given Fe atom, with any spin
on the other Fe atom is the same, i.e. $s_{11} \cdot s_{21} = s_{1i} \cdot s_{21}$ for all $i$, and $J_i=\sum_j J_{ij}$. Keeping this form for the reduced dimer, we
arrive at an Anderson Hamiltonian 
\begin{eqnarray}
H &=& \sum_i J_i s_{1i} \cdot s_{2i} 
  + \sum_{i\sigma} \left[\beta_i (c_{1i\sigma}^\dag c_{2i\sigma} + c_{2i\sigma}^\dag c_{1i\sigma}) + \Delta_i (c_{1i\sigma}^\dag c_{1i\sigma} + c_{2i\sigma}^\dag c_{2i\sigma})\right]
\label{eq:multiorbitalhsup}
\end{eqnarray}
which is the one used in the main text.
Note that when solving for the eigenvalues of the 
Hamiltonian, we restrict each ion to have at most one additional electron 
(i.e. the lowest oxidation state is ferrous).

The above multi-orbital Hamiltonian does not admit an analytic solution. 
However, we can solve for its eigenvalues and eigenvectors numerically.
We have written a code to do this which works with an arbitrary number of base spins and hopping electrons and
which we use also with the [4Fe-4S] Hamiltonian discussed later.
The code is made efficient by working in the basis where the hopping electron is strictly anti-ferromagnetically aligned to the base spins. The hopping matrix elements in this basis can be calculated using Clebsch-Gordan coefficients, similar to Anderson's original work \cite{anderson-DE}. The code can be downloaded with this paper. 

\begin{supptable}
\caption{ The best-fit parameters (cm$^{-1}$)
of the extended Anderson's double exchange Hamiltonian given in Equation~\ref{eq:multiorbitalhsup} used to fit the low-lying energy levels of  the \fereduced dimer. Note that $J_i$ and $\beta_i$ do
{\it not} correspond to the  exchange and double exchange parameters
in the standard HDE model and cannot be directly compared. Standard
deviations (as estimated from the covariance matrix of the fit) given in brackets. \label{tab:multiparams}}
\begin{center}
\begin{tabular}{cccc}
\hline
\hline
&$J_i$&$\beta_i$&$\Delta_i$\\
\hline
1& 2656 ($\pm$513) & 3512 ($\pm$280)&      \\
2& 2743 ($\pm$682) & 9679 ($\pm$294)& 1536 ($\pm$271)\\
3& 2151 ($\pm$518) & 4653 ($\pm$296)& 4433 ($\pm$196)\\
4& 1756 ($\pm$675) & 8472 ($\pm$294)& 6167 ($\pm$268)\\
5& 395  ($\pm$695) & 6562 ($\pm$296)& 6167 ($\pm$284)\\
\hline
\hline
\end{tabular}
\end{center}
\end{supptable}

\begin{suppfigure}
\begin{center}
\resizebox{100mm}{!}{\includegraphics{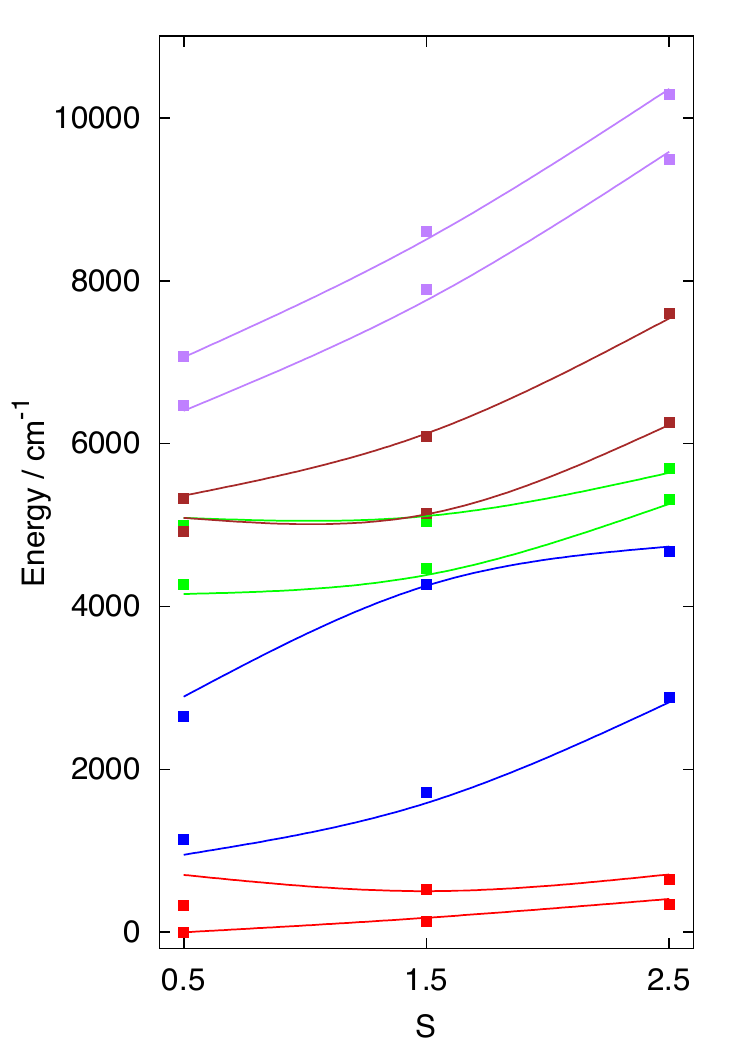}}
\end{center}
\caption{Fit of first 10 states each with spins from S=1/2 to 5/2 using the multi-orbital Anderson model (see Equation~\ref{eq:multiorbitalhsup}).\label{fig:multifit}} 
\end{suppfigure}

A direct fit of the multi-orbital Hamiltonian to the DMRG {\it ab-initio} levels 
yields the parameters  in Supplementary Table \ref{tab:multiparams} and the levels in Supplementary Figure \ref{fig:multifit}. As we can see
the fit is very good; the r.m.s. error is only about 60 cm$^{-1}$. Further it is very robust: out of 12 random initial starting fits, all fits either converged
to the same physical solution shown (to within the standard deviation in the parameters), or attempted to find unphysical solutions with negative parameters.
Overall, this demonstrates that the multi-orbital Hamiltonian indeed captures the essential low-energy physics of the \fereduced complex. Note that the plots in
Fig. \ref{fig:multifit} are for dimer spins $S$=1/2, 3/2, 5/2 only. This is because for
the higher dimer spins, some of the excited states appear to have acquire
d-d transition character, i.e. the Fe ions are not truly high spin. This can be seen, for example in states 8 and 10 for $S$=7/2
in Supplementary Table \ref{tab:3-s2-2}. Such states probably exist in the weak-shoulder region below 10000 cm$^{-1}$ in the low-temperature absorption 
spectrum of ferredoxins \cite{Rawlings01011974}, and cannot be described with the model Hamiltonians we are using. 

As emphasized in the text, the multi-orbital Hamiltonian is {\it not} equivalent to the simple multi-pair generalization of
the HDE model.
This would  correspond to extending the HDE energy levels
in  Eq. (\ref{eq:hdereduced})
to 5 separate pairs of levels arising from each of the pair of d orbitals, 
\begin{equation}
E_i(S)=\Delta_i + 2J_i S_1 \cdot S_2 \pm B_i(S_i+1/2) \label{eq:multipair}
\end{equation}
where the subscript $i$ denotes the pair  involved in the hopping.
This multi-pair HDE model in fact has the same number of parameters as the multi-orbital Hamiltonian (\ref{eq:multiorbitalhsup}) itself. 
However, as seen in the Supplementary Figure \ref{fig:fe3spec} in the main text,  the naive form
does {\it not} fit the {\it ab-initio} DMRG results. Multi-orbital double exchange {\it cannot generally be viewed
simply as the sum of individual orbital double-exchange processes}. In particular,
this means that to be precise we should not characterize double exchange by an effective $B$ parameter as in 
the HDE model, but rather by hopping integrals, $\beta$.

Further support for the multi-orbital nature of the double exchange
is obtained from density difference plots, shown in Supplementary Figure \ref{fig:densitydiff}. 
These plots are obtained as the difference density between different singlet states.
If a single well-defined d orbital pair were to give rise to a pair of states, then
we would expect the density difference to resemble a density associated with a
particular d pair. However, we find that, aside from the lowest two pair of states (which appear to have some $e_g$ parentage), the remaining density differences involve 
contributions from all sets of d orbital densities.

\begin{suppfigure}
\begin{center}
\resizebox{100mm}{!}{\includegraphics{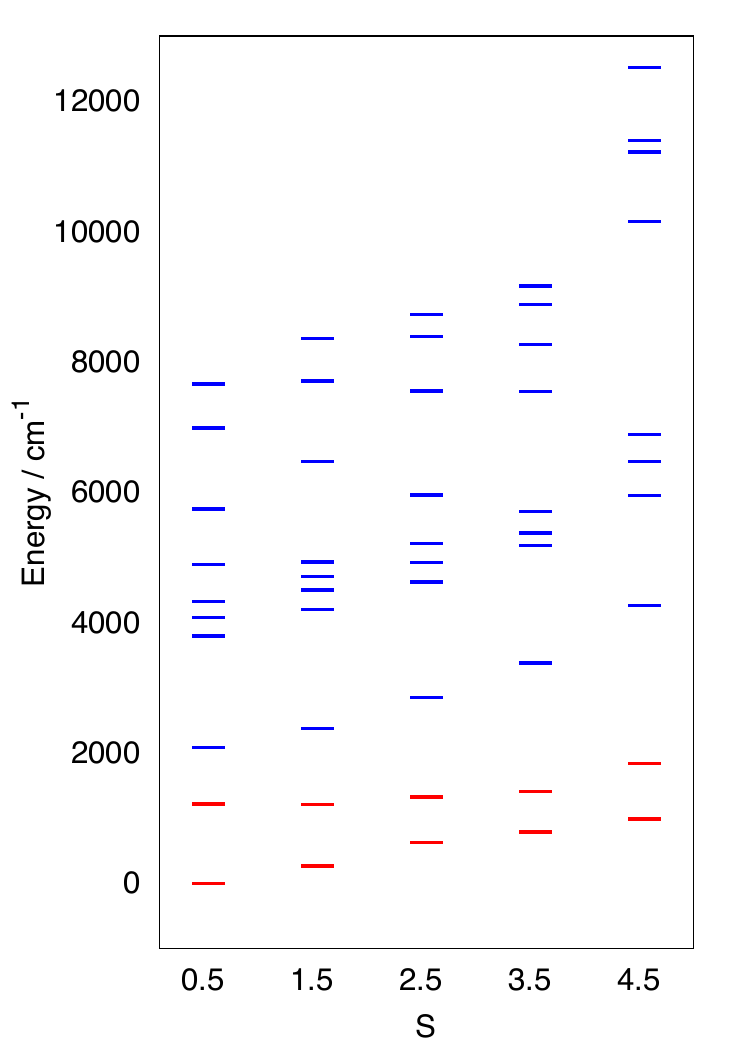}}
\end{center}
\caption{The computed DMRG energy levels at the relaxed \fedimer$^{3-}$ dimer geometry. \label{fig:relaxedfe3levels}}
\end{suppfigure}

We now briefly discuss the effect of geometry relaxation on the \fedimer$^{3-}$ dimer energy levels. At the relaxed geometry,
some localization of the charge occurs. The computed DMRG energy levels at the relaxed geometry are shown in Fig. \ref{fig:relaxedfe3levels}.
We find that our main observations are unchanged: there is little separation between the lowest two and higher energy levels, and the gap between the
lowest two levels does not monotonically increase as required by the HDE model. We have not computed further relaxation effects from solvent
and vibronic coupling, but it is clear from the above that our conclusions about the multi-orbital nature of double exchange
and the need for the multi-orbital Anderson model hold quite generally.

\begin{suppfigure}
\begin{center}
\resizebox{100mm}{!}{\includegraphics{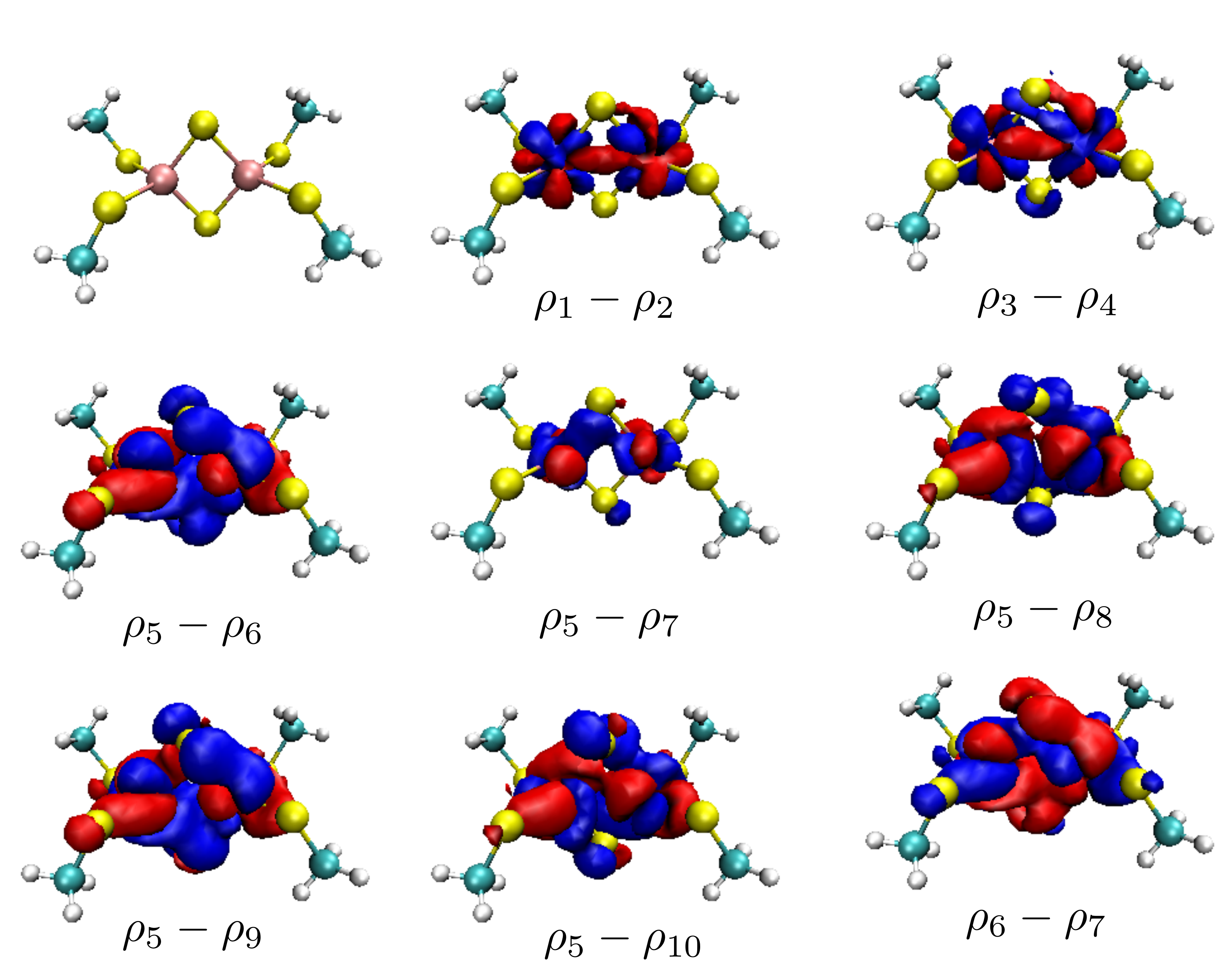}}
\resizebox{100mm}{!}{\includegraphics{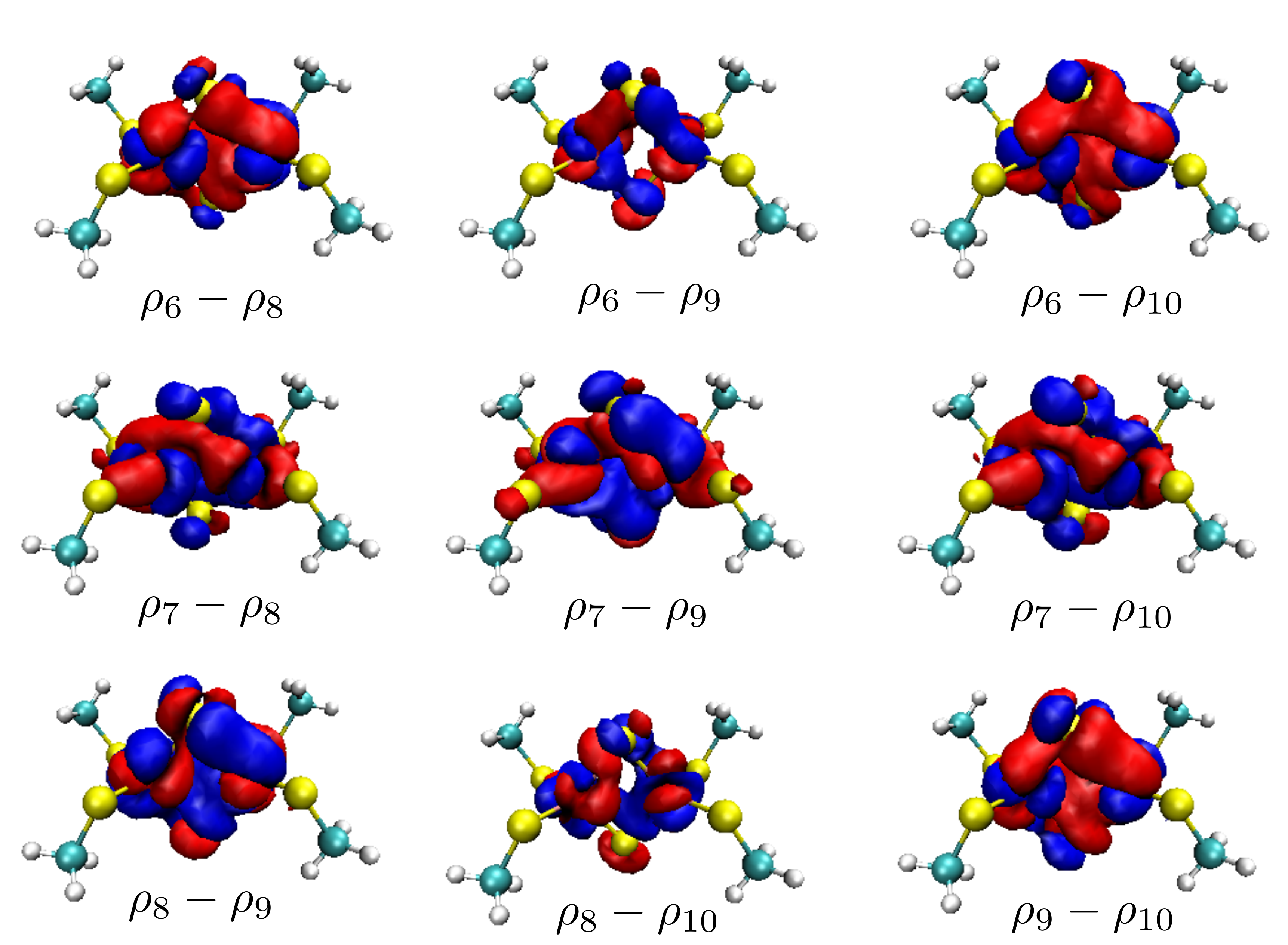}}
\end{center}
\caption{Density differences between sets of doublet states in the
\fedimer$^{3-}$ complex. We notice that besides the lowest two pairs of states, the density difference involves contributions from many d orbitals. \label{fig:densitydiff}}
\end{suppfigure}

\section{4Fe-4S cluster}

\subsection{Geometry and orbitals}

For the [4Fe-4S] cluster, we computed an optimized BS-DFT geometry using the BP86 functional and a triple zeta valence basis (TZV) basis set \cite{tzv-basis}. 
The optimized geometry is shown in Supplementary Table \ref{fig:fecubanegeometry}. 

\begin{supptable}
\caption{Coordinates (in \AA) of the {\fecubane } model complex.}
\label{fig:fecubanegeometry}
\begin{center}
\begin{tabular}{ccccc}
\hline
\hline
& & x & y & z\\
\hline
1& S& 0.04& -1.78& -1.29   \\
2& S& -0.04& 1.78& -1.29   \\
3& S& 1.78& -0.04& 1.29    \\
4& S& -1.78&  0.04& 1.29   \\
5& Fe& 0.05& -1.37& 1.01   \\
6& Fe& -1.38& 0.05& -1.00 \\
7& Fe& -0.05& 1.38& 1.00  \\
8& Fe& 1.37& -0.05& -1.01 \\
9& S&  0.24& 3.30& 2.14   \\
10& S&  -0.24& -3.29& 2.14 \\
11& S&  -3.29& -0.24& -2.14\\
12& S&  3.29& 0.24& -2.14  \\ 
13& C&  -3.80& -1.84& -1.38\\ 
14& H&  -3.91& -1.71& -0.29\\ 
15& H&  -4.76& -2.17& -1.81\\ 
16& H&  -3.03& -2.60& -1.56\\ 
17& C&  3.80& 1.83& -1.38\\
18& H&  3.91& 1.71& -0.29\\
19& H&  4.76& 2.16& -1.81\\
20& H&  3.03& 2.59& -1.55\\
21& C&  -1.83& -3.80& 1.38\\
22& H&  -2.16& -4.76& 1.81\\
23& H&  -2.59& -3.03& 1.55\\
24& H&  -1.70& -3.91& 0.29\\
25& C&  1.84& 3.80& 1.38\\ 
26& H&  2.17& 4.76& 1.81\\ 
27& H&  2.60& 3.03& 1.56\\ 
28& H&  1.71& 3.91& 0.29\\ 
\hline
\hline
\end{tabular}
\end{center}
\end{supptable}

The active space orbitals are chosen in the same way as in the case of the [2Fe-2S] dimer. An unrestricted DFT BP86/SVP calculation was performed at the optimized geometry for the neutral [4Fe-4S] (all ferric) cluster in the high spin ($S_z$=10) state. The  occupied and unoccupied alpha orbitals were then separately localized using the Pipek-Mezey localization technique. The 20 Fe 3d orbitals, 12 bridging S 3p orbitals, as well as 4 terminal ligand S 3p orbitals that point towards the Fe atoms, were
identified by visual inspection. The occupancy of these orbitals in the {\fecubane } cluster gives  an active space of (54e, 36o). Some representative orbitals in the active space are shown in Supplementary Figure \ref{fig:fecubaneorbs}.

\begin{suppfigure}
\begin{center}
\resizebox{160mm}{!}{\includegraphics{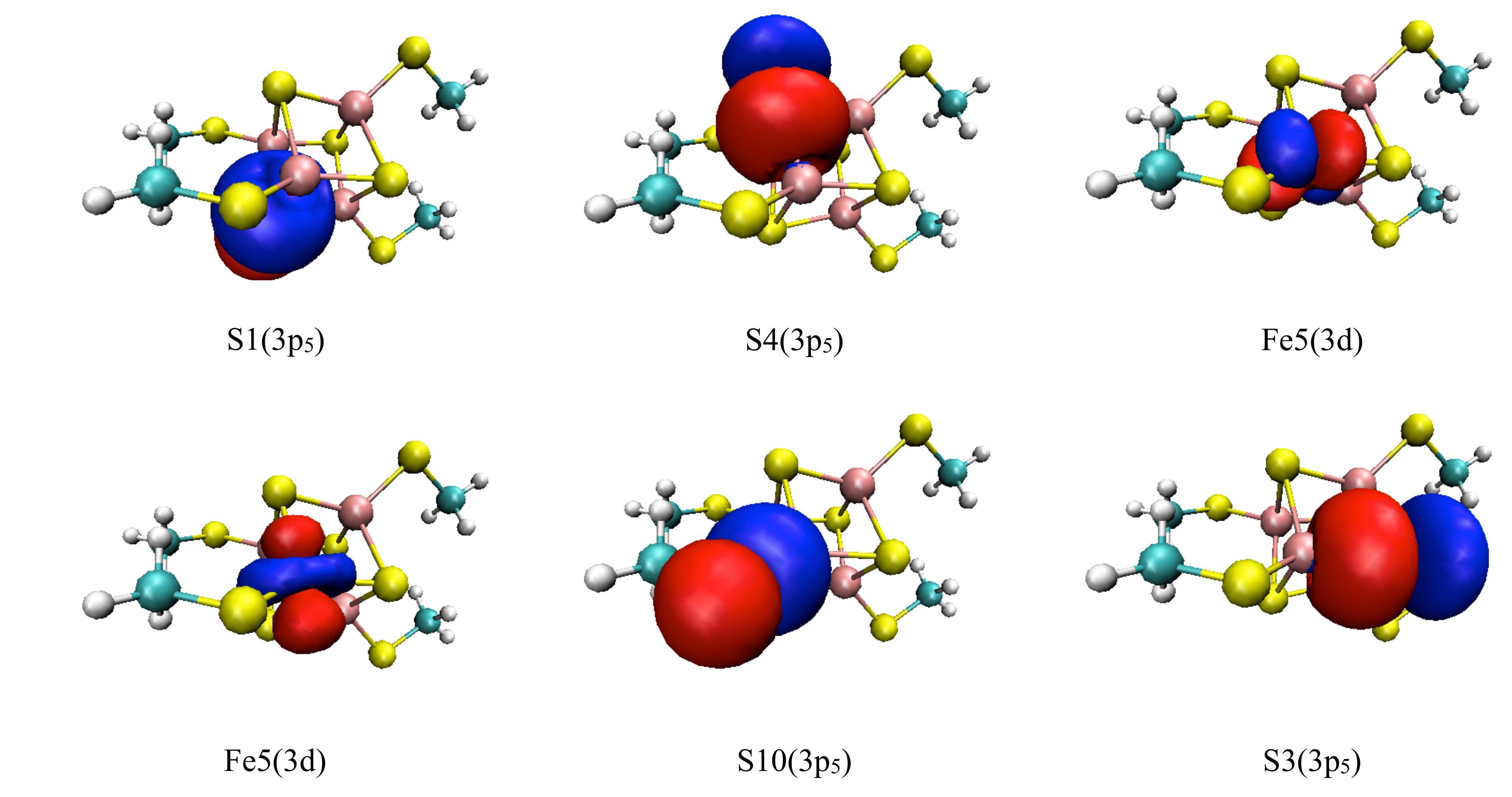}}
\end{center}
\caption{Orbitals in the active space of the {\fecubane } cluster.}
\label{fig:fecubaneorbs}
\end{suppfigure}

\subsection{DMRG calculations}

\subsubsection{Active spaces}
The DMRG-CI calculation with an active space of (54e, 36o) is performed with a maximum of $M$=7500 spin-adapted renormalized states. The 36 orbitals include 20 Fe 3d orbitals, 12 bridging S 3p orbitals, as well as 4 terminal ligand S 3p orbitals that point towards the Fe atoms. For rapid convergence of DMRG energy these orbitals were were ordered as follows: 
S4(3p$_5$), S10(3p$_5$), Fe5(3d), Fe5(3d), Fe5(3d), Fe5(3d), Fe5(3d), S3(3p$_5$), S1(3p$_5$), 
S1(3p$_6$), S4(3p$_6$), Fe6(3d), Fe6(3d), Fe6(3d), Fe6(3d), Fe6(3d), S11(3p$_6$), S2(3p$_6$), 
S4(3p$_7$), S2(3p$_7$), Fe7(3d), Fe7(3d), Fe7(3d), Fe7(3d), Fe7(3d), S3(3p$_7$), S9(3p$_7$), 
S3(3p$_8$), S1(3p$_8$), Fe8(3d), Fe8(3d), Fe8(3d), Fe8(3d), Fe8(3d), S2(3p$_8$), S12(3p$_8$), where the atom labels correspond to the labels in Supplementary Tables~\ref{fig:fecubanegeometry}, and Supplementary Figure~\ref{fig:geom}, and the subscript on S 3p orbitals is the index of the atom they are pointing towards.

\subsubsection{Energy convergence}
As mentioned in the main text of the article (see panel A of Supplementary Figure~\ref{fig:fe4pairs}), in a perfect cubane cluster there are three equivalent pairings of the spins of 
the four Fe atoms to form a singlet ground state. In practice, structural distortion lifts this degeneracy, but the electronic distortion energy is quite small (associated
with an energy scale of less than  8m$E_h$ as seen in \ref{fig:fe4singlettriplet}).
This small energy difference can make robust convergence of the DMRG wave function to the true ground state difficult because the wavefunction 
optimization can get stuck early on in the ``wrong'' pairing.
To expedite the convergence of the wave function towards the correct ground state the Hamiltonian is artificially perturbed so that atoms pairs (Fe5, Fe6) and (Fe7, Fe8) 
gain a strong tendency to ferromagnetically align. This is done by artificially increasing the exchange integrals between the 3d orbitals of the paired up Fe atoms by 0.01 $E_h$ until the number of renormalized states $M$=1600 is reached. Subsequently this perturbation in the Hamiltonian is decreased to zero over the next few sweeps, and 
the rest of the DMRG calculation (up to $M$=7500) is performed on the unperturbed Hamiltonian.

In fact, the above method of perturbing the Hamiltonian can be used to ``converge'' the DMRG wavefunction towards any of the three pairings 
shown in Supplementary Figure~\ref{fig:fe4pairs}. The ground-state pairing is identified as the one with the lowest energy. The other pairings do not necessarily approximate exact 
eigenstates of the full Hamiltonian, but rather, are local minima in the parameter space of DMRG. They are a form of ``broken-symmetry'' DMRG solution, and can be thought of
as the best DMRG states that can be obtained with a maximum $M$=7500, when the spin couplings of the various Fe centers is constrained to be non-optimal. 
The energy differences between these different solutions can be used to
estimate  the difference in the exchange coupling coefficients of the 
HDE model hamiltonian (see Section~\ref{sec:fe4hdeham}).

The DMRG energies and discarded weights at different values of $M$ are used to 
extrapolate to zero discarded weight, which also gives us the estimated energy errors. 
This is shown in Supplementary Figure~\ref{fig:fe4singlettriplet}. Our estimated error in the total energies is less than 1 m$E_h$.

%% This method of guiding the DMRG wave function towards a preferred spin coupling configuration is used to calculate three singlet states and the triplet state. A DMRG calculation was also performed to calculate the energy of the high spin state (S=9), but in this state the local spins of all the Fe atoms are ferromagnetically aligned and no perturbing potential was required to guide the wave function towards the appropriate ground state.

\begin{suppfigure}
\begin{center}
\resizebox{80mm}{!}{\includegraphics{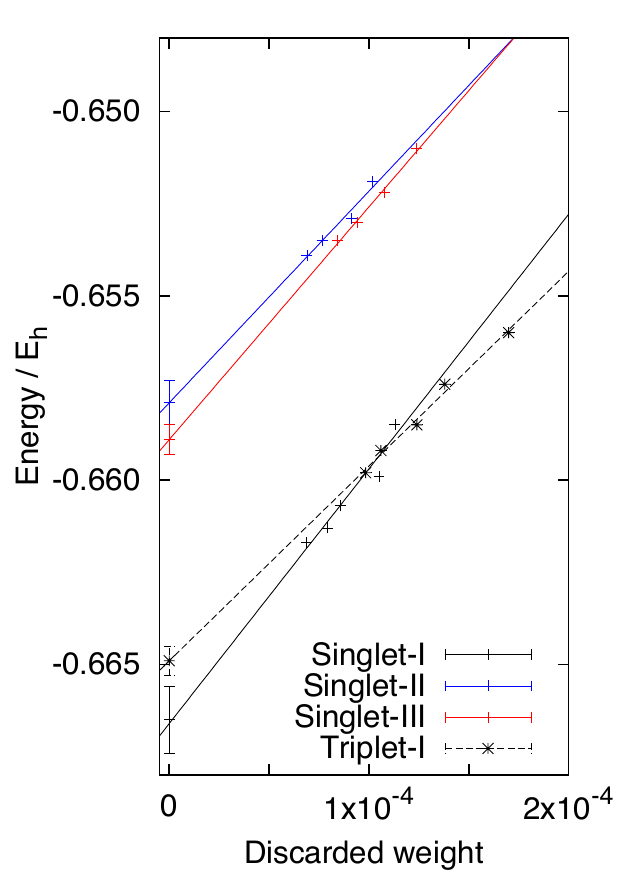}}
\end{center}
\caption{DMRG energy (E+8471.0) in $E_h$ of the ground-state singlet (Singlet-I) and triplet (Triplet-I) states versus the discarded weight of the [Fe$_4$S$_4$(SCH$_3$)$_4$]$^{-2}$ cluster. In addition, the energies of Singlet-II and Singlet-III correspond to the states where the spin pairing is constrained be non-optimal (see text).
The best fit straight lines are extrapolated to zero discarded weight, with the shown error bars, to obtain an estimated FCI energy.}
\label{fig:fe4singlettriplet}
\end{suppfigure}

\subsection{Local charge and spin}
Local populations and spin correlation functions between different Fe atoms 1-4 can be calculated using the equations~\ref{eq:localpop} and \ref{eq:spincorr} and are given in Supplementary Table~\ref{tab:popspincorr}.

\begin{supptable}
\caption{Local population and spin correlation function for the four Fe atoms in the calculated DMRG states of the [Fe$_4$S$_4$(SCH$_3$)$_4$]$^{-2}$ cluster.}\label{tab:popspincorr}
\begin{center}
\begin{tabular}{cccccc}
\hline
\hline
\multirow{2}{*}{Fe atom}&\multirow{2}{*}{ $\langle N_i\rangle$} & \multicolumn{4}{c}{$\langle S_i \cdot S_j\rangle$}\\
&&Fe1&Fe2&Fe3&Fe4\\
\hline
\multicolumn{6}{l}{Singlet-I}\\
\hline
Fe1&6.27&	5.27&	3.24&	-4.05&	-4.05	\\
Fe2	&6.27&	3.24	&5.26	&-4.05&	-4.04	\\
Fe3	&6.27&	-4.05&	-4.05&	5.27&	3.24	\\
Fe4	&6.27&	-4.05&	-4.04&	3.24	&5.27	\\
\hline
\multicolumn{6}{l}{Singlet-II}\\
\hline
Fe1&	6.25&	 5.32&	-4.03&	3.30&	-4.18	\\
Fe2&	6.26	&     -4.03&	5.32	&    -4.18&	3.30	\\
Fe3&	6.26	&     3.30&	-4.18&	5.32	& -4.03	\\
Fe4&	6.25& 	-4.18&	3.30	&    -4.03&	5.32	\\
\hline
\multicolumn{6}{l}{Singlet-III}\\
\hline
Fe1&	6.25&	5.32&	-4.06&	-4.18&	3.31	\\
Fe2&	6.25	&-4.06	&5.33	&3.32	&-4.17	\\
Fe3	&6.25	&-4.18	&3.32	&5.33	&-4.06	\\
Fe4	&6.25	&3.31	&-4.17	&-4.06	&5.32	\\
\hline
\multicolumn{6}{l}{Triplet}\\
\hline
Fe1&	6.26&	5.29&	3.22&	-3.77&	-3.77	\\
Fe2&	6.26	&3.22	&5.27	&-3.95	&-3.95	\\
Fe3&	6.26	&-3.77	&-3.95	&5.28	&3.26	\\
Fe4&	6.26	&-3.77	&-3.95	&3.26	&5.28	\\
\hline
\hline
\end{tabular}
\end{center}
\end{supptable}

\subsection{Model Hamiltonian for pairing and unequal exchange in the [4Fe-4S] cluster} \label{sec:fe4hdeham}

\subsubsection{Single orbital Anderson model}

The direct extrapolation of the simple (single orbital per Fe) Anderson model described in Eq.~(\ref{eq:andersondimer}) to the case of [4Fe-4S] cluster takes the form
\begin{equation}
H = \sum_{ij}J_{ij}(S_i\cdot S_j+S_i \cdot s_j + S_j \cdot s_i)  + \sum_{\sigma={\uparrow,\downarrow}} \beta_{ij}(c^\dag_{i\sigma} c_{j\sigma} + 
c^\dag_{j\sigma} c_{i\sigma})\label{eq:andersonfe4}
\end{equation}
with 6 Heisenberg coupling coefficients $J_{ij}$ and 6 hopping integrals $\beta_{ij}$. We assume that there
are two hopping electrons, and that the two electrons cannot be on the same Fe atom due to on-site repulsion. 

We solve the above Hamiltonian numerically in the space where the spins $s_i$ and $S_i=5/2$ are anti-ferromagnetically aligned. 
%There are
%6 low-lying states, which correspond to the 6 different ways in which the two hopping electrons can be distributed amongst the four ``Fe'' %atoms.  
The nature of the states changes as the ratio of the exchange coupling coefficients 
changes. In Supplementary Figure \ref{fig:hdefe4}
we take $B=2J'$ and  $J_{12}, J_{34}=J'$, $J_{13}, J_{14}, J_{23}, J_{24}=J$. 
For $J/J'>1.2$, we recover a single pairing picture for the $S_{12}$=9/2 dimer states that is assumed in the generalized HDE model
of Noodleman et al. \cite{Noodleman1995,yamaguchi1978extended,Yamaguchi1989210,QUA:QUA21843,yama1998}. 

\begin{suppfigure}
\begin{center}
\resizebox{100mm}{!}{\includegraphics{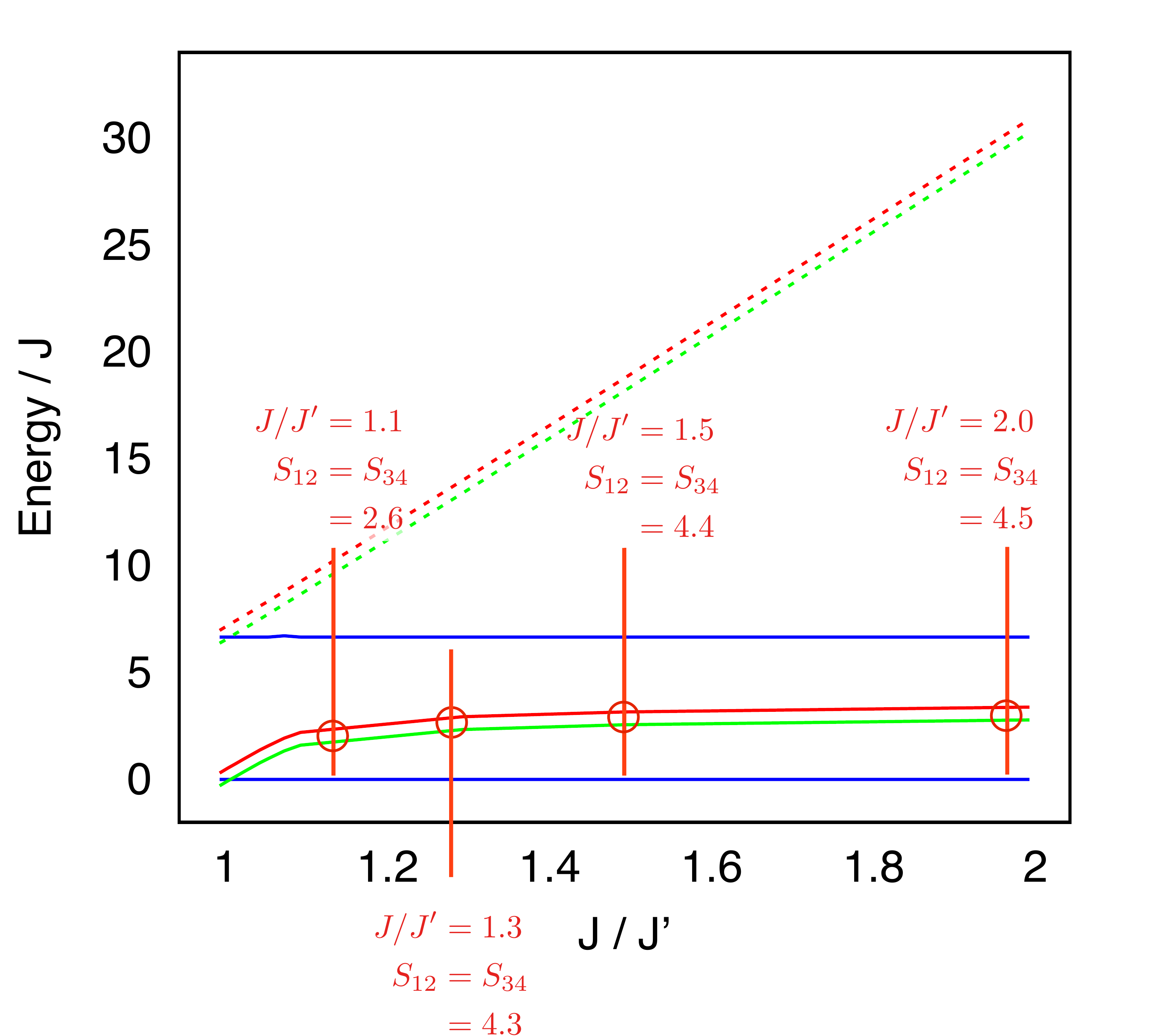}}
\end{center}
\caption{Simplified Anderson model energy levels for the [4Fe-4S] cluster, as a function
of anisotropic exchange $J'\neq J$. As the anisotropy increases, we recover
the fixed-pairing picture assumed by the HDE model. }
\label{fig:hdefe4}
\end{suppfigure}

The difference in the $J_{ij}$ parameters can be estimated from energy differences between the ground state singlet and triplet states and the difference between 
the ground state singlet state (Singlet-I) and artificially paired singlet (Singlet-II). The energy difference between the singlet and the triplet state is given by
\begin{equation}
E(S) - E(T) = J - 0.08|B| \label{eq:stgap}
\end{equation}
and the energy difference between Singlet-I and Singlet-II states is given by
\begin{equation}
E(S_I) - E(S_{II}) = 22.5 (J-J') \label{eq:ssgap}
\end{equation}
In Eq.~\ref{eq:stgap} we assume $B=J$ so that the singlet-triplet gap is $0.92J$. In  Eq.~\ref{eq:ssgap} we have assumed that difference
in the value of $B$ for different couplings is relatively small. Using the converged DMRG energies we obtain $J=382$ cm$^{-1}$ and $J-J' = 84$ cm$^{-1}$, from
which we conclude that  $J/J'\approx 1.28$ and the [4Fe-4S] low-lying states with high effective dimer spin can be described by the single pairing picture, although they
appear to lie close to the border of validity of that description.

\subsubsection{Multi-orbital Anderson model}

To check that the above analysis holds in the more complex case of multi-orbital double exchange, we have also analyzed some multi-orbital (per Fe) Anderson models for the [4Fe-4S] cluster. These take the form
\begin{equation}
H = \sum_{iAB} J_{iAB} s_{iA} \cdot s_{iB} 
  + \sum_{iAB\sigma} \beta_{iAB} (c_{iA\sigma}^\dag c_{iB\sigma} + c_{iB\sigma}^\dag c_{iA\sigma}) +\sum_{iA\sigma} \Delta_{iA} c_{iA\sigma}^\dag c_{iA\sigma}
\end{equation}
where $A$ and $B$ now range over the 4 Fe atoms. The index $i$ ranges over the number of d orbitals on each Fe atom. This would be 5 orbitals in the real cubans, but solving the Hamiltonian for all its levels would be prohibitively expensive. We have therefore considered simpler versions (which illustrate the appropriate trends) where each model Fe atom has respectively only 1 or 2 orbitals. Note that in the 1 orbital case, the maximum spin on each Fe is then only 1/2, and the maximum dimer spin is also 1/2, while
in the 2 orbital case, the maximum spin on each Fe is 1 and the maximum dimer spin is 3/2. (The 1 orbital model is  related to the Hubbard model on a tetrahedron
as discussed in  Refs. \cite{falicov1984exact}).
In Supplementary Figures \ref{fig:fe4spin2} and \ref{fig:fe4spin1} we plot the effective dimer spins of the energy levels as a function of the 
exchange couping ratio (inequivalent $J$'s), for $\beta_{iAB}=2J'$ and $\Delta_{iA}=0$ (for all $i$, $A$). We see the same general trends as in the simple Anderson model above. Note that $J_c$ shifts to lower values as the number of orbitals on each Fe increases.

\begin{suppfigure}
\begin{center}
\resizebox{100mm}{!}{\includegraphics{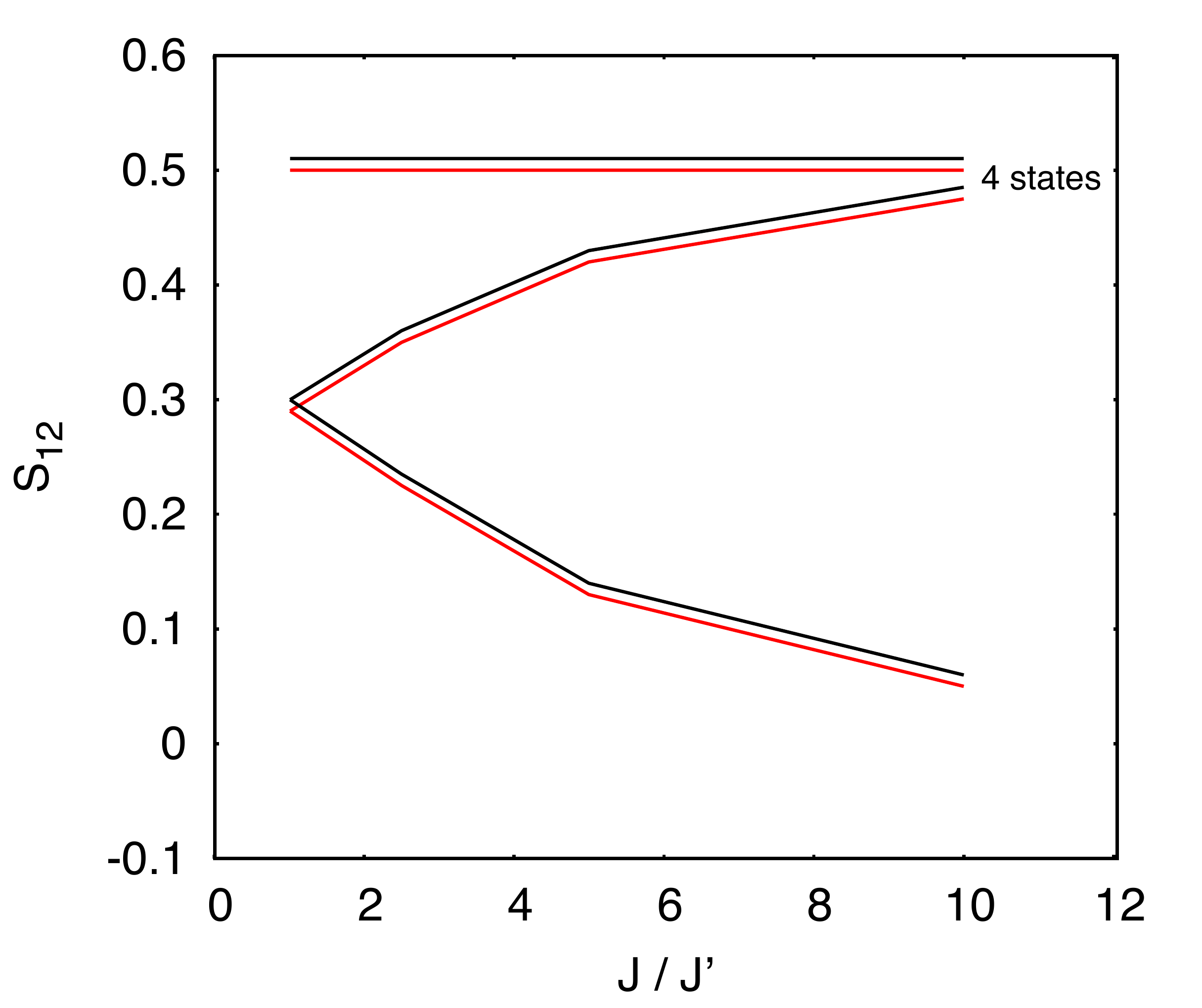}}
\end{center}
\caption{Effective dimer spins of the energy levels as a function of exchange coupling ratio for the [4Fe-4S] multi-orbital Anderson model with 1 orbital per model Fe atom. Note
that since the ``Fe'' atom has only one orbital, the maximum spin per atom is 1/2, and the maximum dimer spin is 1/2. }
\label{fig:fe4spin2}
\end{suppfigure}

\begin{suppfigure}
\begin{center}
\resizebox{100mm}{!}{\includegraphics{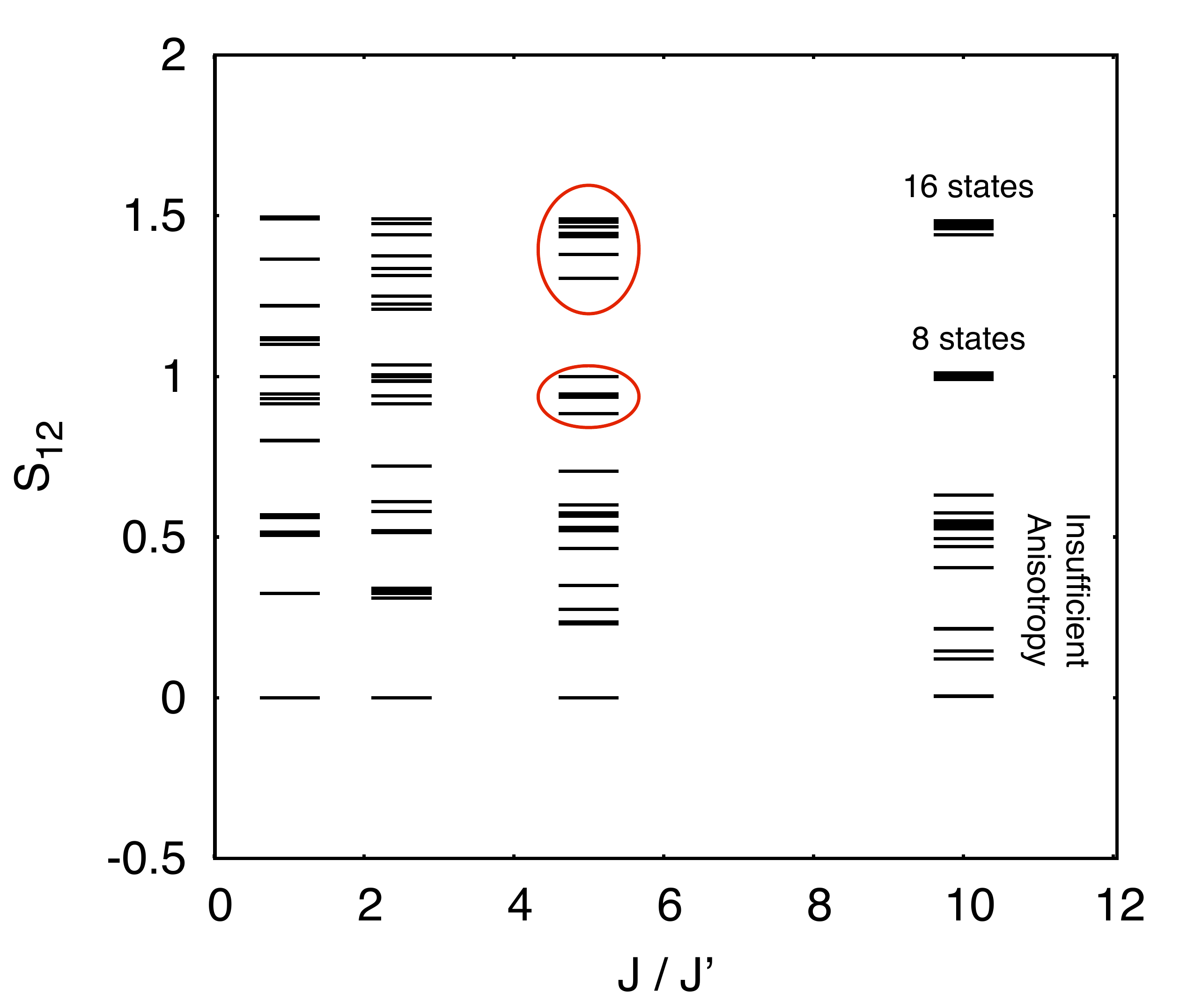}}
\end{center}
\caption{Effective dimer spins of the energy levels as a function of exchange coupling ratio for the [4Fe-4S] multi-orbital Anderson model with 2 orbitals per Fe.
Note that since the ``Fe'' atom has only two orbitals, the maximum spin per atom is 1, and the maximum dimer spin is 3/2.}
\label{fig:fe4spin1}
\end{suppfigure}

%\bibliography{Fe-S}
%\bibliographystyle{Science}